\documentclass[12pt,preprint]{aastex}
\usepackage{epsfig}
\usepackage{lscape}
\usepackage{rotating}
\def\lsim{\lower0.6ex\vbox{\hbox{$ \buildrel{\textstyle
<}\over{\sim}\ $}}}
\def\rsim{\lower0.6ex\vbox{\hbox{$ \buildrel{\textstyle
  >}\over{\sim}\ $}}}
\def\alwaysmath#1{{\ifmmode{#1}\else{$#1$}\fi}}
\def\he#1{\hbox{\alwaysmath{{}^{#1}}{\rm He}}}

\def\li#1{\hbox{\alwaysmath{{}^{#1}}{\rm Li}}}

\def\beq{\begin{equation}}
\def\eeq{\end{equation}}

\def\etal{{\it et al.}~}

\defcitealias{os01}{Paper I}
\defcitealias{os04}{Paper II}

\shortauthors{Aver}
\shorttitle{Non-Parametric Helium Abundances}

\begin{document}

\title{
A New Approach to Systematic Uncertainties and Self-Consistency in Helium Abundance Determinations
}

\author{Erik Aver}
\affil{Physics Department, 
University of Minnesota, Minneapolis, MN 55455}
\email{aver@physics.umn.edu}

\author{Keith~A.~Olive}
\affil{William I. Fine Theoretical Physics Institute, \\
University of Minnesota, Minneapolis, MN 55455}
\email{olive@umn.edu}

\author{Evan D. Skillman}
\affil{Astronomy Department, University of Minnesota,
      Minneapolis, MN 55455}
\email{skillman@astro.umn.edu}

\begin{abstract}
\vskip -6in
\begin{flushright}UMN-TH-2838/10\\FTPI-MINN-10/07\\
January 2010\end{flushright}
\vskip 5in

Tests of big bang nucleosynthesis and early universe cosmology require precision measurements 
for helium abundance determinations.  However, efforts to determine the primordial helium abundance 
via observations of metal poor H~II regions have been limited by significant uncertainties 
(compared with the value inferred from BBN theory using the CMB determined value of the baryon density).  
This work builds upon previous work by providing an updated and extended program in evaluating these 
uncertainties.  Procedural consistency is achieved by integrating 
the hydrogen based reddening correction with the helium based abundance 
calculation, i.e., all physical parameters are solved for simultaneously.  
We include new atomic data for helium recombination and 
collisional emission based upon recent work by Porter \etal and wavelength 
dependent corrections to underlying absorption are investigated.  
The set of physical parameters has been expanded here to include the effects of 
neutral hydrogen collisional emission.  
It is noted that H$\gamma$ and H$\delta$ allow better 
isolation of the collisional effects from the reddening.  
Because of a degeneracy between the solutions for density and temperature, 
the precision of the helium abundance determinations is limited.  
Also, at lower temperatures (T $\lesssim$ 13,000 K) the neutral hydrogen fraction is poorly 
constrained resulting in a larger uncertainty in the helium abundances.  Thus the derived errors 
on the helium abundances for individual objects are larger than those typical of previous studies.  
Seven previously analyzed, ``high quality'' H~II regions are utilized for the primordial helium 
abundance determination.  The updated emissivities and neutral hydrogen correction generally raise 
the abundance.  From a regression to zero metallicity, we find Y$_{p}$ as 0.2561 $\pm$ 0.0108, in 
broad agreement with the WMAP result.  Tests with synthetic data show a potential 
for distinct improvement, via removal of underlying absorption, using higher resolution 
spectra.  A small bias in the abundance determination can be reduced significantly
and the calculated helium abundance error can be reduced by $\sim$ 25\%.  

\end{abstract}


\newpage
\baselineskip=18pt
\noindent

\section{Introduction}

Big bang nucleosynthesis (BBN) provides us with a unique window to the Universe at redshifts of order $10^{10}$ \citep{wssok,osw,fs}.  With the WMAP determination of the baryon density of the universe \citep{wmap}, BBN has become a zero-parameter theory \citep{cfo2,cfo3}, leading to relatively precise predictions of the light element abundances of D, $^{3}$He, $^{4}$He, and $^{7}$Li \citep{cfo,coc,coc2,cyburt,coc3,cuoco,serpico,cfo5}.   In addition to lending itself to scrutiny by comparing the predictions of BBN with observations, the resulting abundances are of significant scientific value in constraining the content and interactions of the universe during BBN \citep{MM93,sar,cfos}.  
These constraints, however, generally require accurate determinations of the light element
abundances, particularly in the case of \he4 where better than 1\% determinations are necessary.
The difficulties in obtaining this precision for \he4 is the focus of this paper.

The concordance between BBN and observations of the CMB temperature anisotropies
is predominantly due to the agreement between the BBN predicted primordial value of 
D/H compared with determinations made from observations of high redshift quasar absorption 
systems.  The cosmic microwave background (CMB) power spectrum provides a high precision measurement of the baryon density or equivalently the baryon-to-photon ratio, 
$\eta$, (primarily through the first and second acoustic peak heights).  
Using the 5-year WMAP value of $\Omega_B h^2 = 0.02273 \pm 0.00062$ or
$\eta = (6.23 \pm 0.17) \times 10^{-10}$ \citep{wmap}, one expects D/H $= (2.49 \pm 0.17) \times 10^{-5}$ \citep{cfo5} which should be compared with $(2.82 \pm 0.21) \times 10^{-5}$  \citep[][and references therein]{kirk03, pettini}.
On the other hand,  $^{7}$Li is predicted to be substantially higher \citep{cfo5,hos}
than determinations of \li7 abundances in the outer layers of halo dwarf stars.  This places extra emphasis on the determination of the primordial helium abundance, Y$_{p}$.  As a result of the very weak logarithmic relationship between $\eta$ and Y$_{p}$, the measured error on Y$_{p}$ must be very small ($<1\%$) to achieve a reasonable constraint on the determination of $\eta$.  For comparison, the WMAP value of $\eta$ enables one to predict the value of Y$_{p}$ to within 0.08\%.

Being relatively chemically unevolved,  low metallicity H~II regions in dwarf galaxies can be used to provide a measure of Y$_{p}$.  
Since metallicity only increases with age, these galaxies can be used to fit the helium abundance versus metallicity and therefore 
allows one to extrapolate back to very low values of O/H, used here as a surrogate for metallicity, corresponding to the primordial 
helium abundance \citep{ptp74}.  However, after more than three decades of work in determining the abundance using this method, 
the measurements frequently have not agreed with each other and until recently, post-WMAP, have not been in accordance with the WMAP 
value (see Figure \ref{Yp_t} below), $Y_p = 0.2486 \pm 0.0002$, which is based on the BBN calculation of \citet{cfo5} assuming the WMAP determined value of $\eta$.

The quoted errors in the extragalactic determination of Y$_{p}$ have been historically very small in comparison to the range in the results (again see Figure \ref{Yp_t}).  The early determinations of Y$_p$ focused on statistical errors in the measured helium line fluxes and, as a result, underestimated the error.  Ideally, the target quantity, the helium to hydrogen ratio (y$^{+}={n(He~II) \over n(H~II)}$) in extragalactic H~II regions, would directly follow from an observed helium to hydrogen emission line ratio.  However, as a result of interstellar reddening, underlying stellar absorption, optical depth, and collisional corrections, the actual determination of y$^{+}$ is complicated, and the true uncertainty can be large.  Interstellar reddening occurs as the photons are scattered by dust on their journey.  Being a wavelength dependent process, each line is affected differently.  The stellar continuum juxtaposes absorption features under nebular emission lines.  The H~II region itself can absorb some of the emitted photons as well.  Finally, the total emission fluxes are due to both recombination and collisional excitation.  In the case of the hydrogen emission, only the small fraction of recently recombined neutral atoms can contribute to collisional emission.  Unfortunately, these confounding processes can not be directly measured.  In addition, the temperature, which determines the physical rates, cannot be determined directly and independently unless it is assumed to be perfectly uniform throughout the H~II region.  Our ignorance of these necessary physical model parameters will necessarily inhibit the determination of y$^{+}$ and therefore correspond to potentially large increases in the error on that determination.  The error on Y$_p$ also varies widely due to sample sizes.  The error on the intercept, Y$_p$, will shrink with increasing number of regression points; however, poor quality spectra do not allow for a robust estimation of the systematic effects listed above.  Therefore, this work focuses on a small number of ``high quality'' extragalactic objects.

\begin{figure}
\plotone{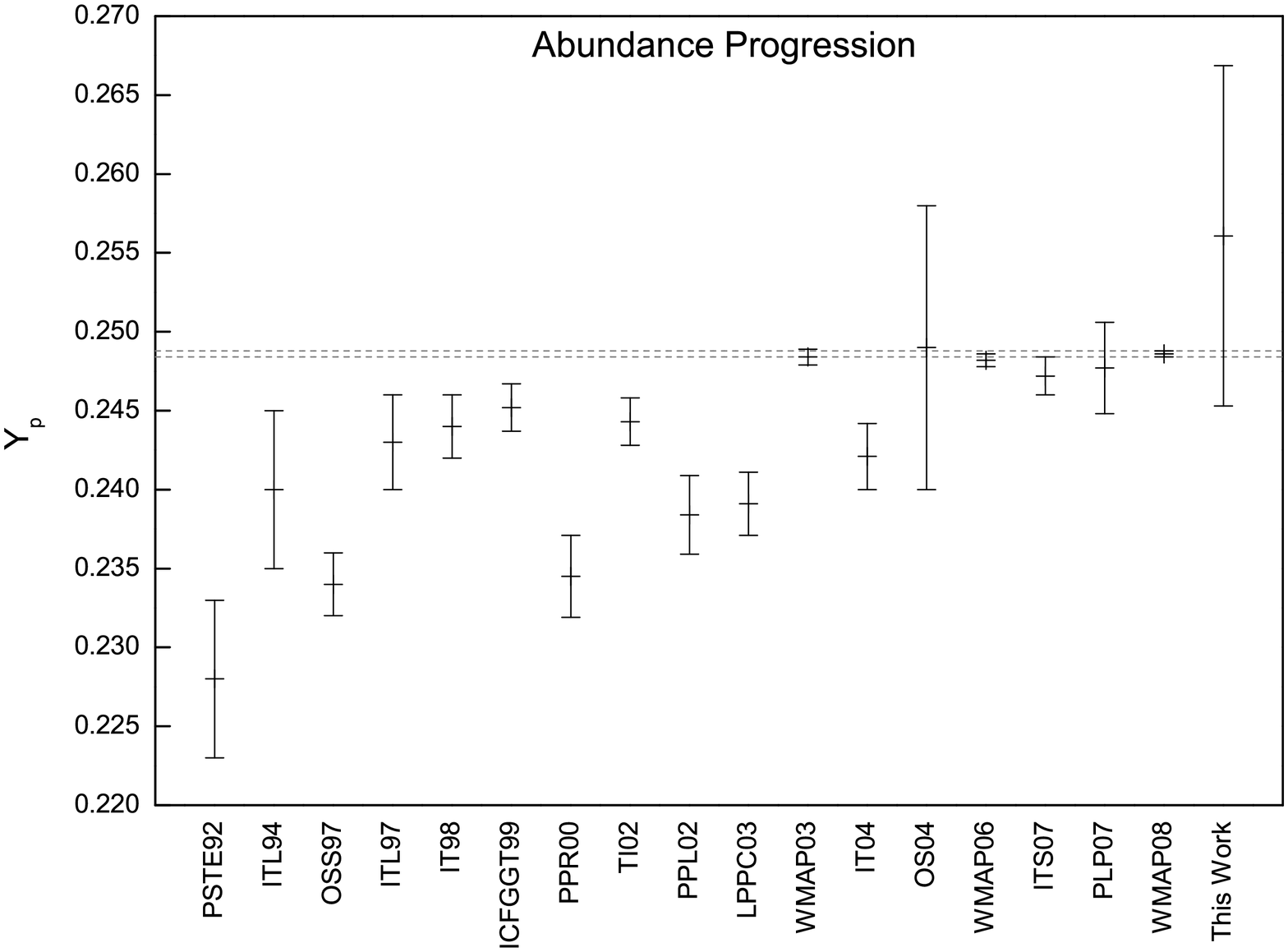}
\caption{
Recent historical progression of the primordial helium mass fraction over the previous decade and comparison to the WMAP result \citep[with the values given in date progression by][labeled as PSTE92,
ITL94, OSS97, ITL97, IT98, ICFGGT99, PPR00, TI02, PPL02, LPPC03, WMAP03,
IT04, OS04, WMAP06, ITS07, PLP07, WMAP08, and AOS09, for this work,
respectively]{pag92,itl94,oli97,itl97,it98,izo99,ppr00,ti02,ppl02,lur03,spe03,it04,os04,spe06,its07,plp07,wmap}
.  A general increase in the primordial abundance is apparent.  Note that, historically,
the error bars are typically small relative to the differences between studies and with WMAP in particular.
}
\label{Yp_t}
\end{figure}

Previous work by \citeauthor{os01} in \citetalias{os01} \citeyearpar{os01} 
and \citetalias{os04} \citeyearpar{os04}, following the ``self-consistent'' 
methods of \citet{itl94} in determining the electron density (used for the 
collisional correction) and \citet{ppr00} in determining the  
electron temperature (used for the helium emissivities), addressed some of the  
systematic effects by 
solving for physical conditions via a minimization of the 
differences between the values of y$^{+}$ calculated from each 
of the helium lines.  The $\chi^{2}$ minimization is accomplished by a 
down-hill simplex algorithm.  
Monte Carlo simulations over the input fluxes were used to measure the
statistical variance with the ultimate effect of further increasing the total 
uncertainty.  Details of the determination of the model parameters and the 
impact of the Monte Carlo process on these parameters will be discussed 
further in section \ref{Determining y} below.

The purpose of the current work is to explore further refinements to the 
approach of Papers I \& II.  First, there now exist improvements to the 
atomic data \citep[][PFM]{pfm07} that can be incorporated 
\citep[e.g.,][hereafter ITS07 \& PLP07 respectively]{its07,plp07}.  
Second, previously, the flux ratios of the hydrogen lines were used to
solve for reddening and underlying hydrogen absorption, while the flux
ratios of the helium lines were used to solve for density, temperature,
underlying helium absorption, helium self-absorption, and the helium abundance.
There is no need to solve for these two sets of parameters separately, 
and doing so artificially suppresses any potential degeneracies between them.
Thus, here we explore the effects of combining two separate ``self-consistent'' 
approaches into one.  
Third, the previous work assumed identical values of equivalent width of 
underlying absorption for all H~I or He~I emission lines, regardless of wavelength,
and here we investigate the effects of a  wavelength dependence in the 
underlying stellar absorption (e.g., ITS07).
Fourth, we study the effects of correcting for the collisional emission 
from neutral hydrogen  \citep[e.g.,][]{dk85,sk93,si01,lur03}.  The second 
change, the integration of the hydrogen and helium, is a calculational 
procedure modification, while the other three are improvements in the 
modeling of the physical processes that impact the observed hydrogen and 
helium emission lines.  Detailing the incorporation of these four effects 
and investigation of their impact is the primary aim of this paper and 
is discussed in sections \ref{Determining y}-\ref{NHCC}.  
In section \ref{icfs}, we discuss the importance of ionization correction 
factors.  Section 
\ref{Results} presents a new determination of the primordial helium abundance
resulting from the reprocessing of the observations studied in \citetalias{os04} 
with comparison to those results, and the results reported in ITS07 
and PLP07. Section \ref{Discussion} details the effect of the enhancements, 
remaining errors, and possibilities for future improvements.

\section{Determining y$^{+}$} \label{Determining y}

As described in \citetalias{os01}, six measured helium emission line fluxes ($\lambda$3889, 4026, 4471, 5876, 6678, and 7065) relative to $H\beta$ ($\frac{F(\lambda)}{F(H\beta)}$) and their equivalent widths ($W(\lambda)$) are used to determine the helium abundance.  Corrections are made for reddening, underlying absorption, collisional enhancement, and radiative transfer:
\beq
y^{+} = \frac{F(\lambda)}{F(H\beta)}\frac{E(H\beta)}{E(\lambda)}{\frac{W(\lambda)+a_{He}}{W(\lambda)} \over \frac{W(H\beta)+a_{H}}{W(H\beta)}}\frac{1}{f_{\tau}(\lambda)}\frac{1}{1+\frac{C}{R}(\lambda)}10^{f(\lambda)C(H\beta)},
\eeq
where $\frac{E(H\beta)}{E(\lambda)}$ is the emissivity ratio of H$\beta$ to the helium line.  In the approach of papers I and II, the reddening relative to H$\beta$ (C(H$\beta$)) and underlying hydrogen absorption (a$_{H}$) are determined separately by Monte Carlo over the three hydrogen line ratios ($\frac{H\alpha}{H\beta}$, $\frac{H\gamma}{H\beta}$, and $\frac{H\delta}{H\beta}$).\footnote{The reddening coefficients, f($\lambda$), are calculated from the extinction fits of \citet{ccm89}.}  The optical depth function \textit{$f_{\tau}$}, and collisional to recombination emission ratio \textit{$\frac{C}{R}$} are both temperature and density dependent (the emissivities are also temperature dependent).  Therefore the physical parameters needed for the corrections are electron density (n$_{e}$), underlying helium absorption (a$_{He}$, corrected for in terms of equivalent width), optical depth ($\tau$), and temperature (T).  The parameters are determined simultaneously with the abundance by minimizing $\chi^{2}$, defined as the weighted difference between each helium line's abundance and the average.  This minimization is performed for each realization in a Monte Carlo over the input helium line fluxes
and equivalent widths.

An important result of the Monte Carlo analysis was the discovery of degeneracies in the sense that the fit parameters were not always independently constrained \citepalias{os01,os04}.  In particular, the temperature and density exhibited a trade-off whereby increasing the temperature while decreasing the density would leave the $\chi^{2}$ relatively unchanged.  The reason for this can be seen in the form of the emissivities (see \S \ref{Updated Emissivities}).  Four of the lines show similar exponential dependencies on temperature and density. As a result, the abundance from these lines shifts correspondingly for changes in temperature and density.

\begin{figure}
\plotone{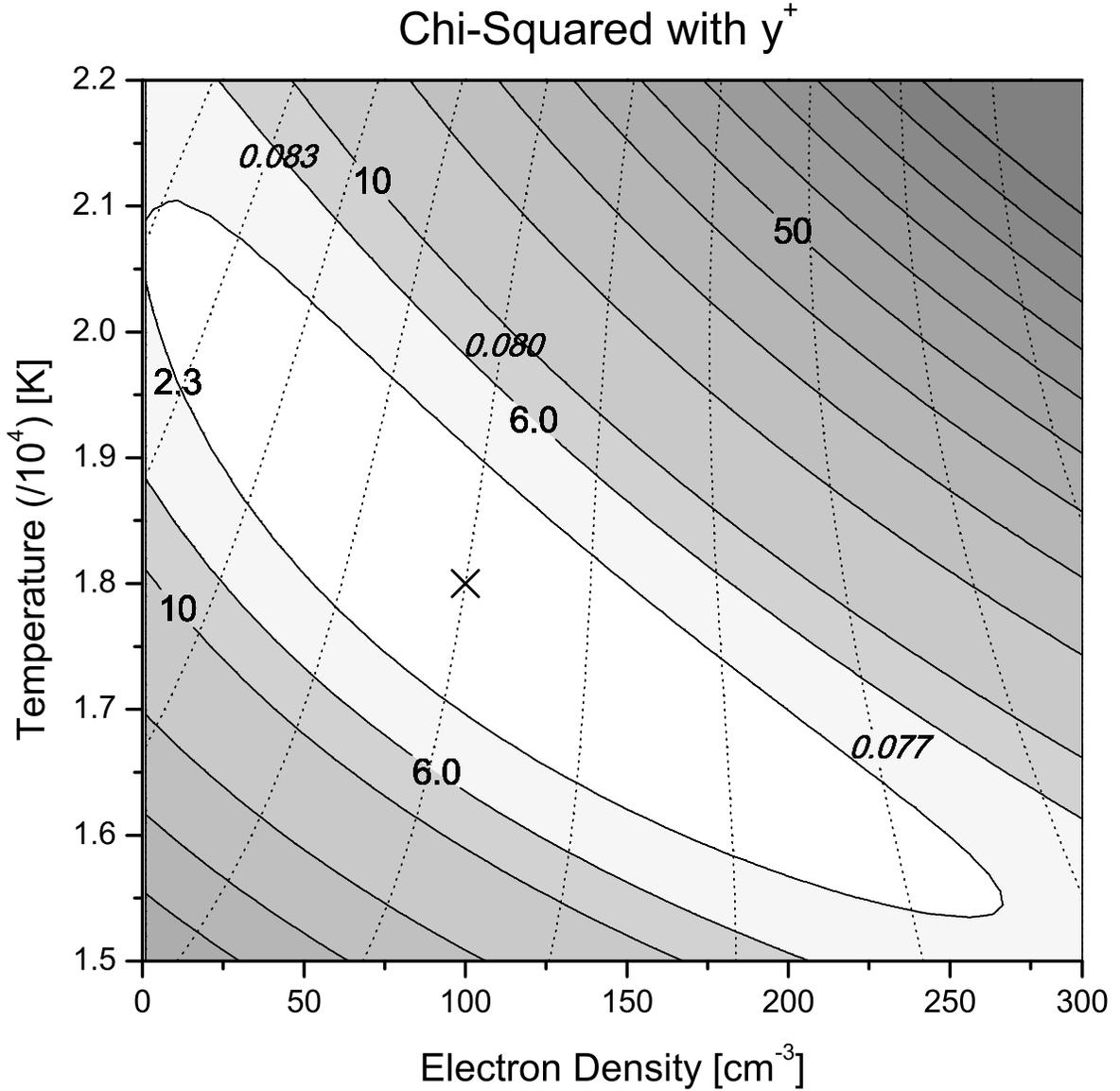}
\caption{
A plot of the derived helium abundance as a function of the temperature and density for a synthetically generated spectrum.  Impressed on this diagram is a contour plot of $\chi^2$ versus density and temperature for the same spectrum.  The synthetic model uses $n_{e} = 100~cm^{-3}$, $a_{He} = 1.0$ \AA, $\tau = 1.0$, $T = 18,000$ K, and $y^{+}=0.08$.  The extension of the $\chi^{2} = 2.3$ contour with a strong negative correlation highlights the degeneracy between density and temperature (note that for synthetic data $\chi^{2}_{min} = 0.0$).  That the $\chi^{2}$ and abundance contours are nearly perpendicular demonstrates the impact of the degeneracy on the abundance determination ($\pm$5\%).}
\label{X2}
\end{figure}

To demonstrate the importance of the degeneracy between temperature and density, characteristic model parameters and a helium abundance were chosen and used to generate synthetic flux data.  $\chi^{2}$ was then calculated keeping the abundance and all of the traditional fit parameters fixed except for the density and temperature.  The impact of trading the density against the temperature can be clearly seen in Figure \ref{X2} which illustrates a well constrained quotient of density and temperature but a $\triangle\chi^{2} = 2.3$ boundary spanning a temperature range of 5500 K, a density range of 275 cm$^{-3}$, and an abundance variation of 10\%.  The shallow, extended minimum results in a large range of best fit parameters for the density and temperature upon the Monte Carlo perturbation, and the abundance gains a much larger uncertainty (see Figure \ref{Y_D_MC} below).  Only by solving for these two parameters simultaneously can this degeneracy be discovered.  

\begin{figure}
\plotone{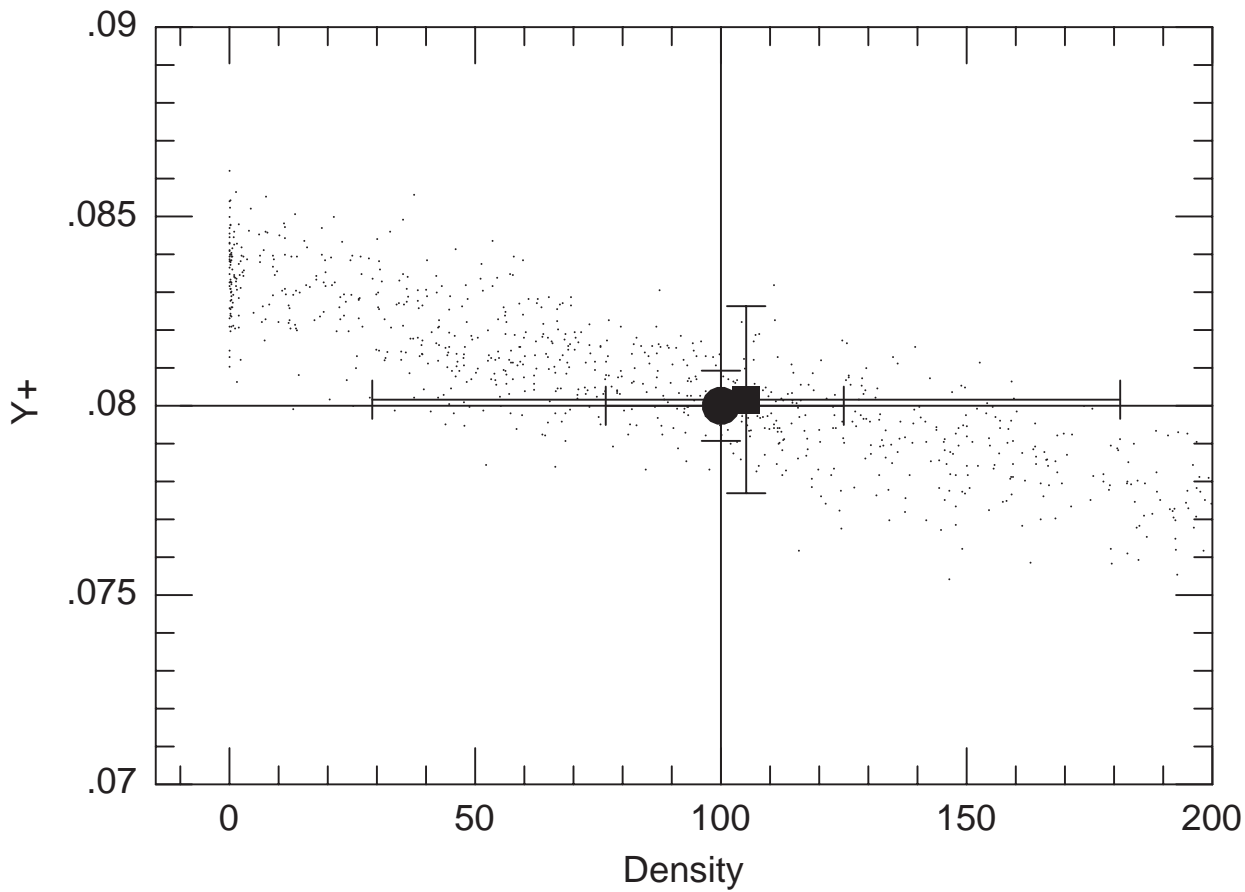}
\caption{
Monte Carlo plot of 1000 solutions based upon synthetic data taken from \citetalias{os01}.  The synthetic model is for $n_{e} = 100~cm^{-3}$, $a_{He} = 0.1$ \AA, $\tau = 0.1$, and $T = 18,000$ K.  The solid circle with error bars marks the original (direct) solution; while the solid square with error bars marks the average of the Monte Carlo solutions.  Upon performing the Monte Carlo, the large range in density solutions (with a corresponding large range in temperature solutions), gives a much larger density uncertainty, resulting in a marked increase in the abundance uncertainty (1\% to 3\%).
}
\label{Y_D_MC}
\end{figure}

The inclusion of wavelength dependent absorption, now \textit{a$_{H}(H\beta)$} and \textit{a$_{He}(\lambda)$}, and neutral hydrogen collisional emission, \textit{$\frac{C}{R}(H\beta)$}, will modify the helium to hydrogen ratio as,
\beq
y^{+} = \frac{F(\lambda)}{F(H\beta)}\frac{E(H\beta)}{E(\lambda)}{\frac{W(\lambda)+a_{He}(\lambda)}{W(\lambda)} \over \frac{W(H\beta)+a_{H}(H\beta)}{W(H\beta)}}\frac{1}{f_{\tau}(\lambda)}\frac{1+\frac{C}{R}(H\beta)}{1+\frac{C}{R}(\lambda)}10^{f(\lambda)C(H\beta)}.
\eeq
The wavelength dependent absorption is included in order to model variations in the amount of stellar absorption and is represented by a relative coefficient for each line.  The neutral hydrogen collisional emission is included to account for the small population of non-ionized hydrogen which can therefore undergo collisional emission.  This process is composed of the temperature dependent collisional excitation and the cascade of downward transitions and requires the introduction of a new parameter, the neutral to ionized hydrogen density fraction, $\xi$.  In \citetalias{os04}, this was assumed to be negligible.  Primarily, the neutral hydrogen collisional emission corrects the hydrogen lines but through H$\beta$ shows up in the abundance.  The integration of the hydrogen reddening and helium abundance calculation is in principle accomplished by simply summing their $\chi^{2}$ contributions and minimizing over the now expanded set of physical parameters (n$_{e}$, a$_{He}$, $\tau$, T, C(H$\beta$), a$_{H}$, $\xi$) simultaneously.  The combined $\chi^{2}$ is given by,
\beq
\chi^2 = \sum_\lambda {(X_R(\lambda) - X_T(\lambda))^2 \over  \sigma(\lambda)^2} + \sum_{\lambda} {(y^+(\lambda) - {\bar y})^2 \over \sigma(\lambda)^2}
\label{eq:X2}
\eeq
where,
\beq
X_R(\lambda) = \frac{F(\lambda)}{F(H\beta)}{\frac{W(\lambda)+a_{H}(\lambda)}{W(\lambda)} \over \frac{W(H\beta)+a_{H}(H\beta)}{W(H\beta)}}\frac{1+\frac{C}{R}(H\beta)}{1+\frac{C}{R}(H\lambda)}10^{f(\lambda)C(H\beta)}
\eeq
\beq
\bar y = \sum_{\lambda} {y^+(\lambda) \over \sigma(\lambda)^2} / \sum_{\lambda} {1 \over \sigma(\lambda)^2}.
\eeq
X$_{R}$ represents the corrected hydrogen flux, X$_{T}$ the theoretical hydrogen emission, $\bar y$ the average helium abundance, and $\sigma$ is calculated from the measured statistical error and equivalent width error.  

In our analysis below, we use the same dataset of H~II regions as in \citetalias{os04} for calculating y$^{+}$ and extracting Y$_{p}$.  Those galaxies were chosen from the \citet[][IT98]{it98} sample with screening to minimize uncertainties in underlying absorption, 
to select only low O/H systems, and to avoid contamination by Galactic Na~I absorption.  In addition to those seven targets, NGC~346A \citep{ppr00}, which was also examined in \citetalias{os04}, 
and I~Zw~18 (ITS07) are analyzed.  This allows for a comparison with each of the five targets used in PLP07.  All nine except NGC~346 are included in the recent large sample of ITS07.  The spectrum for Mrk~193 originally reported in \citet{itl94} and used in \cite{os04} has been reanalyzed and the results of the new analyses are reported in \cite{its07}.  Here we use the newly analyzed spectrum for Mrk~193.  

Each object is analyzed with the new corrections applied successively to the procedure used in \citetalias{os04}.  The analysis is conducted with the parameter space restricted to physically meaningful values, and Monte Carlo average values and standard deviations are returned for the model parameters and abundance.  Reflecting a straightforward improvement in the atomic data and modeling, first the updated emissivities are used in conjunction with separate hydrogen reddening and helium abundance calculations.  Next, the hydrogen and helium calculations are combined into one integrated calculation.  The third change is to introduce the slight wavelength dependence in the underlying absorption.  Finally, the correction for neutral hydrogen collisional emission is incorporated.  In the following sections, each of these enhancements is first discussed theoretically and then applied to the dataset.

\section{Updated Emissivities} \label{Updated Emissivities}

 PFM have recently provided improved helium emissivity fits, as a function of temperature, and collisional contributions, as a function of temperature and density, based upon the most recent atomic data \citep[see][]{por05}.  They report broad agreement between these emissivities and their plasma simulation code CLOUDY \citep{fer98} of better than 0.03\% within the temperature range $5000 \leq T \leq 25,000 K$.  However, the abundance determination also requires the H$\beta$ emissivity.  Though not in the scope of the published paper, R. L. Porter (private communication), provided us  with the H$\beta$ emissivity in terms of the published paper's parameterization,
\beq
E(H\beta) = [-2.6584\times10^5 - 1420.9~(\ln T)^2 + 35,546~\ln T + \frac{6.5669\times10^5}{\ln T}]~T^{-1}.
\eeq

\begin{figure}
\resizebox{0.9\textwidth}{!}{\includegraphics{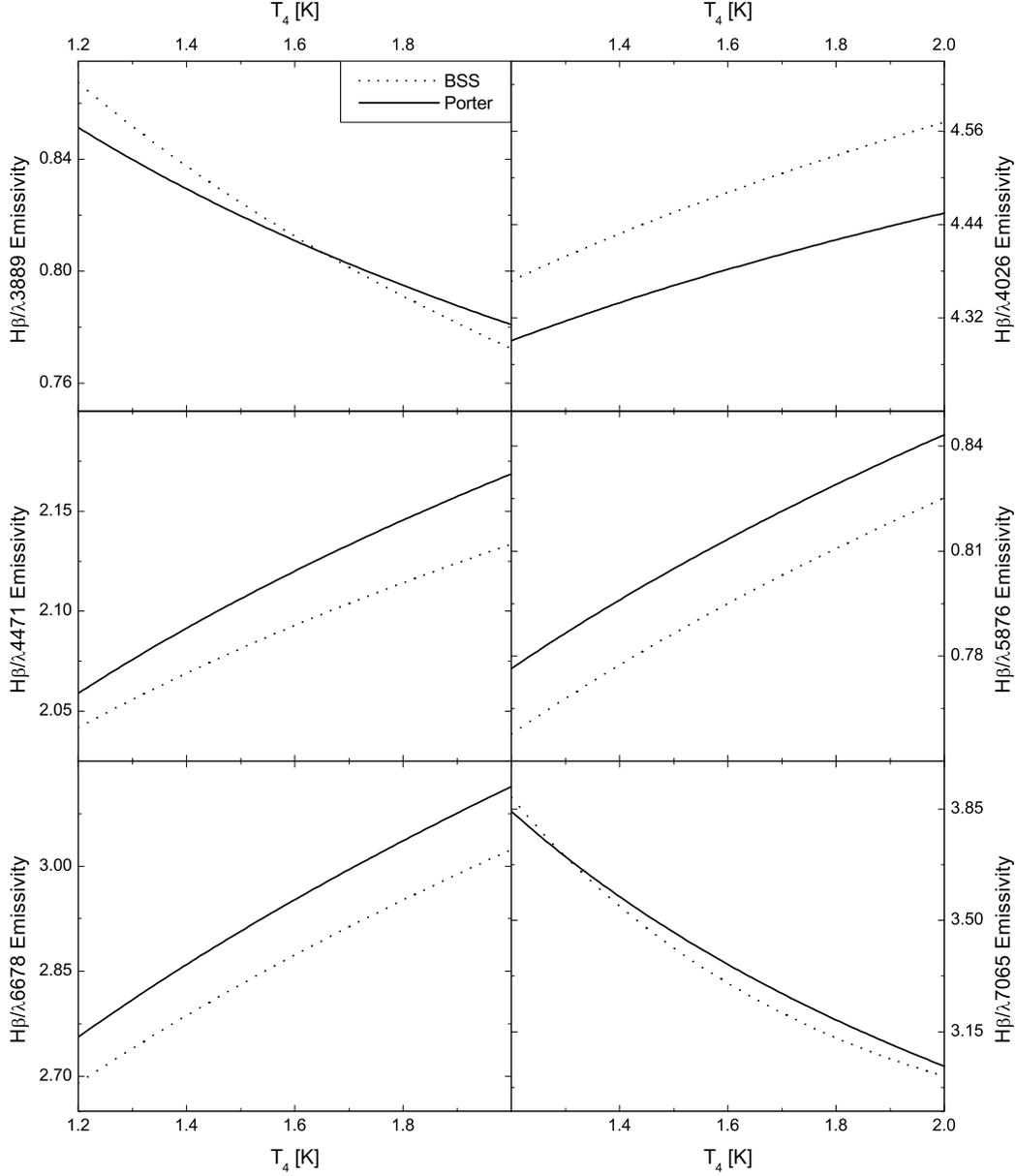}}
\caption{
Comparison of the PFM emissivities, $\frac{E(H\beta)}{E(\lambda)}$, to those of BSS (at n$_e$ = 100 cm$^{-3}$).  The BSS fits are the dashed lines while the PFM fits are solid.  The progression is, left to right, top to bottom, by wavelength: $\lambda$3889, 4026, 4471, 5876, 6678, 7065.  The PFM emissivities plotted here are the refit equations of equation \ref{eq:UE}; on this plot, the equations reported in PFM are within the line thickness of the refit equations.
}
\label{Emissivity}
\end{figure}

\begin{figure}
\resizebox{\textwidth}{!}{\includegraphics{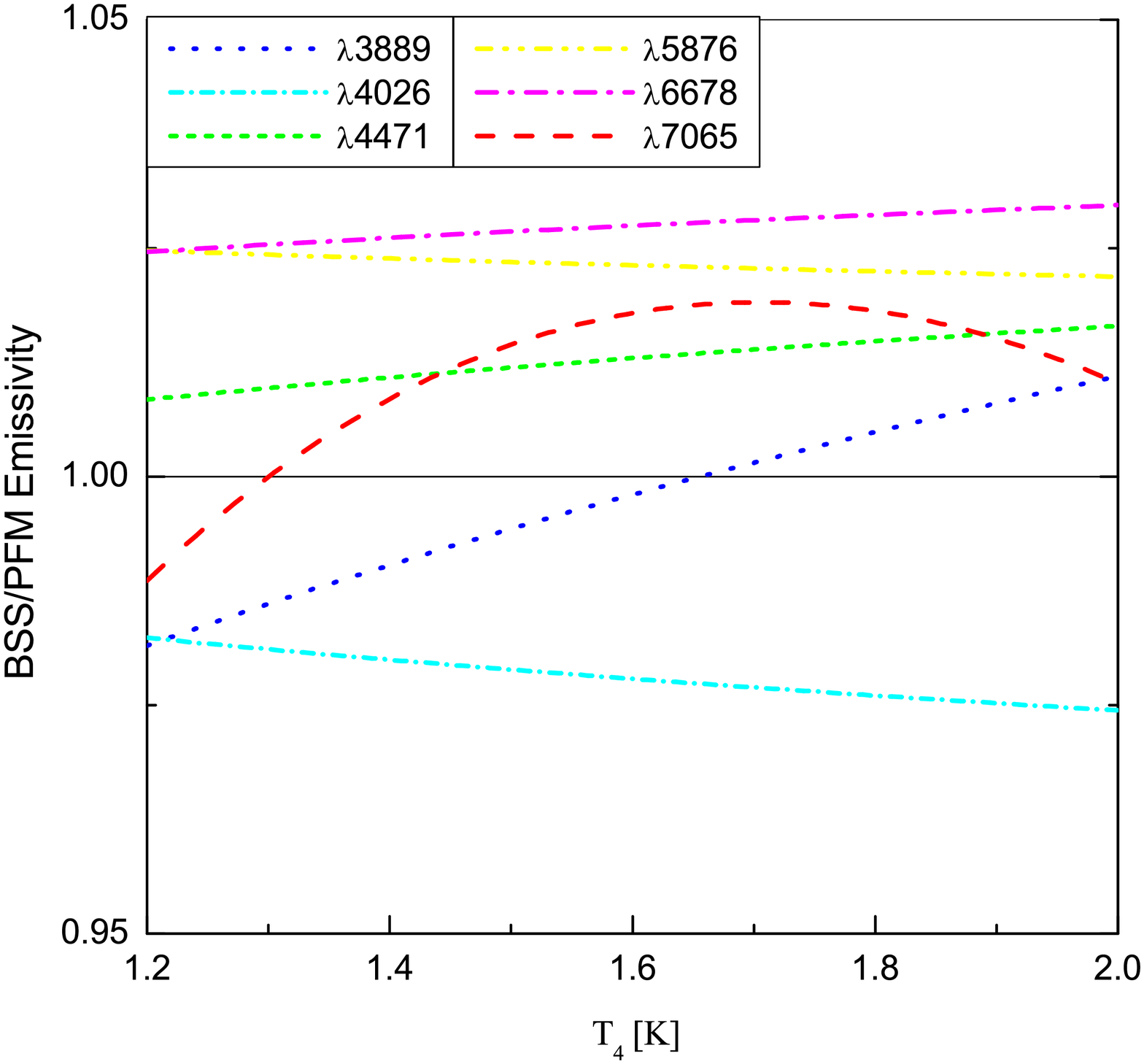}}
\caption{
The PFM emissivities, $\frac{E(H\beta)}{E(\lambda)}$, plotted relative to those of BSS (at n$_e$ = 100 cm$^{-3}$).  The newer fits deviate from the old by only several percent, but the relative shifts are clearly not the same for all six lines.  The three strongest lines, $\lambda$4471, 5876, and 6678, show similar behavior, but 4026 is opposite, and 3889 and 7065 cross the zero deviation line.
}
\label{EmissivityRelative}
\end{figure}

In implementing these new emissivities the following difficulty was encountered during $\chi^{2}$ minimization. As parameterized by PFM, the density dependence of the emissivities weakens with increasing density, ultimately becoming negligible.  While this occurs at densities which are unreasonably high for our objects, the addition of the temperature-density degeneracy can and does lead to densities that increase nearly without bound.  In practice, the Monte Carlo realizations were often found to tend to higher densities that were unconstrained by the form of the emissivities (in a region beyond their applicability).  The resulting $\chi^{2}$ for these solutions decreased only marginally with the large increase in density; so the result is effectively an equivalent solution but which is physically meaningless.  Upon calculating the Monte Carlo average, even a small number ($<1\%$) bias the returned values, especially the density, to an unacceptable degree.  The density dependence in the optical depth function disfavors ever increasing density but only if the optical depth itself does not approach zero.  Refitting the emissivities, with the collisional correction, to the form of \citet[][BSS]{bss99} avoids the pitfall while sacrificing less than 1\% accuracy in the region of interest, $12,000 \leq T \leq 20,000 K$ and $1 \leq n_e \leq 300~cm^{-3}$, and compares favorably with the BSS fit accuracy.\footnote{This is different from the approach taken in ITS07 where the BSS emissivities are divided by a linear temperature dependent fit to the ratio of the BSS and PFM emissivities.}  Therefore, within the physically applicable region for our objects, the emissivity values themselves are nearly identical to those of PFM.  At large densities however, the exponential dependence on the density relegates ever increasing densities to ever increasing $\chi^{2}$.  In effect, the re-parameterized fit protects the $\chi^{2}$ minimization from unphysical solutions while retaining the numerical accuracy in the physical region.  Figures \ref{Emissivity} and \ref{EmissivityRelative} below compare the BSS fits used in \citetalias{os04} with the new fits. The re-parameterized helium emissivities, including the collisional correction and scaled to H$\beta$, are,

\begin{eqnarray}
F_{\lambda} ~~~& = & \frac{E(H\beta)}{E(\lambda)}\frac{1}{1+\frac{C}{R}(\lambda)} \\*
F_{3889} & = & 0.8779 T^{-0.128 - 0.00041 n_e} \nonumber \\*
F_{4026} & = & 4.233 T^{0.085 - 0.00012 n_e} \nonumber \\*
F_{4471} & = & 2.021 T^{0.121 - 0.00020 n_e} \nonumber \\*
F_{5876} & = & 0.754 T^{0.212 - 0.00051 n_e} \nonumber \\*
F_{6678} & = & 2.639 T^{0.244 - 0.00054 n_e} \nonumber \\*
F_{7065} & = & 5.903 T^{-0.519} /(1.462 -
(0.127-.00076 n_e + 0.000000255 n_e^2)T)
\label{eq:UE}
\end{eqnarray}

\subsection{Hydrogen Emission}

In addition to the updated helium emissivities, the theoretical hydrogen emission line ratios taken from \citet{hs87}, were refit to incorporate the reported (very weak) density dependence.  In principle, this is included for consistency, but became of particular interest in physically describing the impact of higher density so as to disfavor non-physical solutions.  The previous logarithmic form in temperature was retained and employed for the density fit.  The expanded equations are as follows (with Table \ref{table:HEmiss} listing the coefficients),

\beq
X_T = \sum_{ij} c_{ij} (\log(T_4))^i (\log(D))^j  \nonumber \\ 
\eeq

\begin{table}[ht!]
\centering
\caption{Coefficients for the hydrogen emissivities, $c_{ij}$}
\vskip .1in
\begin{tabular}{lcccc}
\hline\hline
\tabletypesize{\footnotesize}
Line                           & 
{i$\downarrow$}			&
{j$\rightarrow$}  & & \\
\hline
& & 0 & 1 & 2 \\
\hline
H$\alpha$	&	0	&	2.87		&      -0.000813	&      -2.05E-5	\\
		&	1	&      -0.510		&	0.00443		&	3.15E-5	\\
		&	2	&	0.345		&      -0.00969		&	5.69E-5	\\
H$\gamma$	&	0	&	0.468		&	7.58E-6		&	2.19E-6	\\
		&	1	&	0.0281		&      -0.000101	&      -4.76E-6	\\
		&	2	&      -0.0106		&	0.000128	&	2.51E-6	\\
H$\delta$	&	0	&	0.259		&	2.09E-9		&	2.56E-6	\\
		&	1	&	0.0207		&      -1.94E-6		&      -6.51E-6	\\
		&	2	&      -0.00770		&	4.58E-5		&	3.48E-6	\\
		\hline
\label{table:HEmiss}
\end{tabular}
\end{table}

\subsection{Tracking the Effects}

\begin{figure}\centering  
\resizebox{\textwidth}{!}{\includegraphics{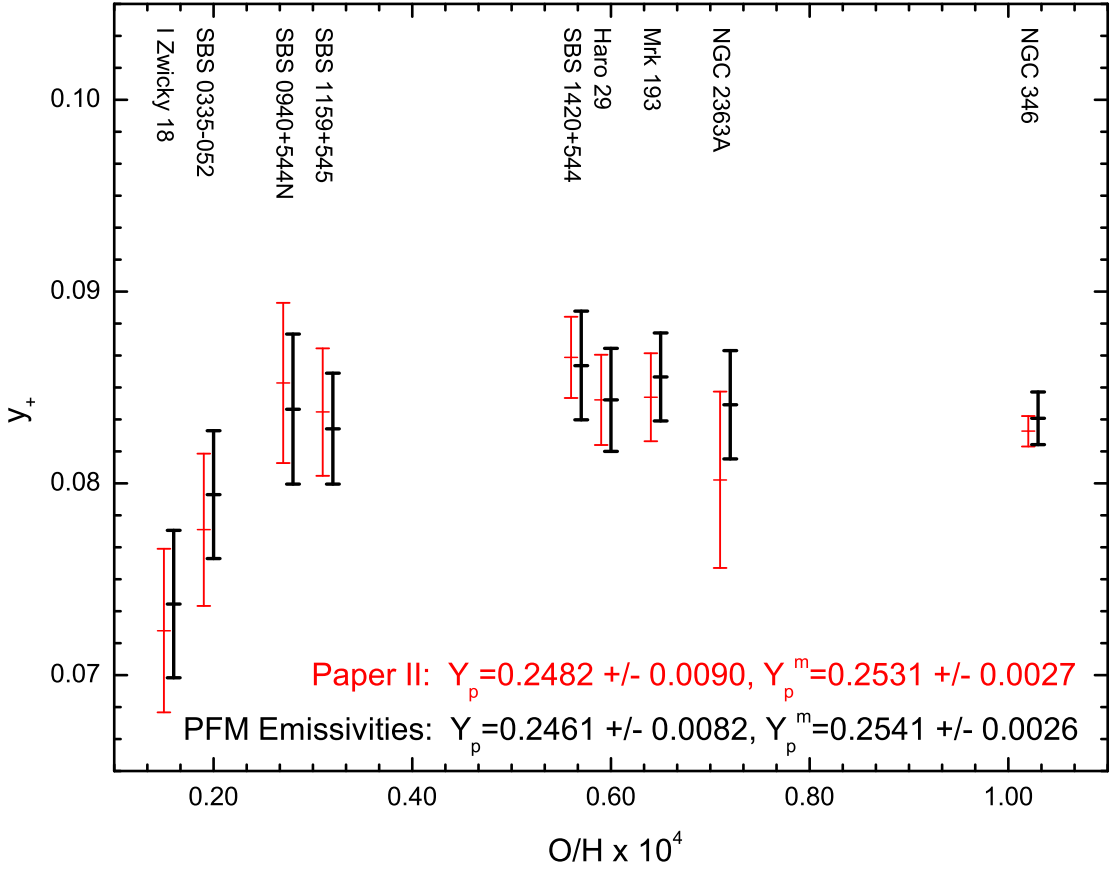}}
\caption{
Abundance comparison for the target objects analyzed as in \citetalias{os04} with the BSS and PFM emissivities.  The red thinner lines are for BSS with PFM given in thicker black bars.  The quoted primordial helium abundance, Y$_{p}$, is based on a regression of the seven objects of \citetalias{os04} (i.e., I~Zw~18 and NGC~346 have not been used in the regression).
}
\label{y_Progression_UE}
\end{figure}

A comparison of the resulting He abundances using BSS and PFM is shown in Figure \ref{y_Progression_UE}.
The updated emissivities, including the collisional correction, in general raise the calculated abundance
of helium.  However, because the changes in different lines are qualitatively different (again see Figures \ref{Emissivity} \& \ref{EmissivityRelative}), the update does not categorically raise the abundance.  For $\lambda$4471, 5876, and 6678, the H$\beta$ to He emissivity is increased rather uniformly (2-3\%) across the relevant temperature range in comparison with those of BSS.  All three lines exhibit a gradual increase with temperature.  The behavior of $\lambda$4026 is similar except that the new emissivity is shifted to lower values.  $\lambda$3889 is more complicated in that the emissivity is reduced at low temperatures but is enhanced at $T \ga$ 16,000 K.  $\lambda$7065 is similar, though the enhancement begins at 13,000 K.  
Broadly, $\lambda$3889, $\lambda$7065, and $\lambda$4471 track the BSS emissivities 
more closely (within $\sim$ 2\%) than the other three lines.  Therefore, upon repeating the \citetalias{os04} analysis with the new emissivities, these lines will favor returning a similar abundance.  The three strongest lines with the smallest relative errors will produce a larger abundance, thus raising the average abundance.  To decrease the discrepancy between the $\lambda$3889 and 7065 and $\lambda$4471, 5876, and 6678 abundances, the temperature will decrease, thus lowering the average abundance slightly.  The weak $\lambda$4026 is shifted opposite to the three strongest lines, but still can have a significant impact as described below.

As anticipated, in most cases, the updated emissivities of PFM raise the abundance 
(see Figure \ref{y_Progression_UE}).  This was observed by PLP07 and ITS07.  
Though, on average, the abundance increases by $\sim$1\%, three of the objects actually 
exhibit slightly decreased helium abundances.  The decreased abundances of 
SBS~940+544N and SBS~1159+545 can be explained by discrepant high $\lambda$4026 fluxes.  
This has been demonstrated using synthetic data, and using the parameters from 
\citetalias{os04} for these two objects, we estimate that $\lambda$4026 fluxes 
may be high by 15\% and 10\% respectively.  Note that the typical uncertainties 
in the equivalent widths of the $\lambda$4026 lines and the corrections for 
underlying absorption are both comparable in magnitude to this suspected
discrepancy.

Overall, in comparison to BSS, the PFM emissivities favor a compromise whereby the temperature decreases and the average abundance increases.  However, as we have seen, this effect is not uniform and an overly strong $\lambda$4026 can cause the abundance to decrease.  
Thus, the regression performed on the seven objects\footnote{I~Zw~18 and  NGC~346 do not pass the three cuts based on equivalent width, oxygen abundance, and radial velocity as discussed in \citetalias{os04}.} of \citetalias{os04} yields a slightly {\em lower} intercept as indicated in Figure \ref{y_Progression_UE}.
It is also worth noting that for all nine objects the temperature decreases as expected.  Since this temperature decrease blunts the abundance increase, calculations based on temperatures not determined simultaneously with the abundance would exhibit a more pronounced abundance increase upon replacing the BSS emissivities with those of PFM.  Qualitatively very similar to BSS, the new emissivities do not dramatically shift the abundances or their pattern.  The more accurate atomic data is a distinct improvement but is not the primary limitation in determining the abundance.

\section{Combined H and He} \label{Integrated}

In Papers I \& II, the spectra are analyzed in a two step approach.  First the hydrogen lines are used in conjunction with the temperature as determined by [O~III] emission to determine the reddening parameter and underlying hydrogen absorption.  These two corrections are then applied to the helium to H$\beta$ input flux ratios.  Subsequently, these corrected lines are used to determine the density, helium absorption, optical depth, and temperature simultaneously with the abundance.  This separation is natural as the dominant hydrogen line correction is for reddening.  This allows the helium spectra to be corrected for interstellar effects before resolving the competing nebular emission and absorption effects on the helium lines.  However, the independence of these calculations is artificially imposed and could impair the abundance calculation.  Since reddening impacts both the hydrogen and helium fluxes (and in the same way), it should be a minimization parameter for both.  Furthermore, even though the hydrogen emissivity ratios are only weakly temperature dependent, the estimated electron temperature is as physically meaningful for the hydrogen emission as for helium.

In order that the calculations be truly self-consistent, the hydrogen and helium based calculations are combined into a single minimization.  This has the advantage of being sensitive to all degeneracies.  Conceptually the implementation of combining the two calculations is straightforward.  Nine line ratios are now input, and all seven physical parameters (n$_{e}$, a$_{He}$, $\tau$, T, C(H$\beta$), a$_{H}$, $\xi$) are determined simultaneously with the abundance via a single minimization.  Equation \ref{eq:X2} explicitly gives this combined $\chi^{2}$; though, for this section the neutral hydrogen collisional correction is not yet included.  The previously observed degeneracies provide a further impetus for integrating the hydrogen line based reddening calculation with the actual abundance calculation.  The larger parameter space allows for new degeneracies but the additional calculational effects could break them.  

An important detail of this process is that the deblending of He($\lambda$3889) and H8 can now be done consistently based upon the solution for reddening and temperature.  Previously, in the sequential analysis, He($\lambda$3889) is deblended using a the hydrogen reddening parameters and a fixed temperature.  As a result, even though the helium lines are used to determine the temperature, the He($\lambda$3889) flux could not be adjusted to account for this temperature change.  This effect is very small since the H8/H$\beta$ emissivity is only very weakly temperature dependent \citep[$\frac{F(H8)}{F(H\beta)} = 0.104~T_{4}^{0.046}$, from a fit to the data of][]{hs87}.  However, integration allows the blended input flux to be deblended during the $\chi^{2}$ calculation using the solved hydrogen underlying absorption, reddening, and temperature.  In this way, the deblending itself affects the $\chi^{2}$ and therefore the relevant parameter solutions.  The $\frac{He(\lambda3889)}{H\beta}$ flux is calculated by correcting the blended input for underlying absorption in H8, reddening the H8 intensity fraction, and subtracting it,
\beq
\frac{F(3889)}{F(H\beta)} = \frac{F(3889+H8)}{F(H\beta)}\frac{W(3889)+a_{H}(H8)}{W(3889)}-0.104\ T_{4}^{0.046}\ 10^{-f(3889)C(H\beta)}
\eeq
(the same fraction is also subtracted from the equivalent width).
The error on this flux is not decreased so as to take the most conservative approach.  Also, the errors on the equivalent widths, not being reported, are taken to be the larger of 10\% or twice the fractional error in the flux.  

\subsection{Investigating the Expanded Parameter Space} \label{Int_Investigation}

Though the theoretical grounds for integrating the helium and hydrogen $\chi^{2}$ minimizations are sound, this guarantees no practical benefit, and, in fact, the expansion of Monte Carlo and parameter space warns of more freedom for non-physical realizations.  Synthetic testing provides some insight since the actual physical parameters are known.  Characteristic values are chosen and used with the model equations to generate corresponding fluxes for all nine line ratios.  For this analysis, n$_{e}$ = 100 cm$^{-3}$, a$_{He}$ = 1.0 \AA, $\tau$ = 1.0, T = 18,000 K, C(H$\beta$) = 0.1, a$_{H}$ = 1.0 \AA, and $\xi$ = 1.0 x 10$^{-4}$, were used in conjunction with EW(H$\beta$) = 250 \AA~and y$^{+}$ = 0.08.  Flux errors of 2\% were assumed.  The generating and solved parameters and abundances are listed in Table \ref{table:Integrated}.

\begin{deluxetable}{lccc}
\tabletypesize{\footnotesize}
\tablewidth{0pt}
\tablecaption{Sequential and Integrated Synthetic Analysis}
\tablehead{
&
\colhead{Input}		 &
\colhead{Sequential}     &
\colhead{Integrated}}
\startdata
He$^+$/H$^+$			& 0.08 	 & 0.08024 $\pm$ 0.00268 & 0.07995 $\pm$ 0.00324 \\
T$_e$				& 18,000 & 18,212 $\pm$ 1768 	 & 18,053 $\pm$ 3039 \\
N$_e$				& 100.0  & 122.5 $\pm$ 123.4 	 & 166.4 $\pm$ 307.9 \\
ABS(He~I)			& 1.0	 & 1.02 $\pm$ 0.08 	 & 1.01 $\pm$ 0.10 \\
$\tau$				& 1.0    & 0.93 $\pm$ 0.40 	 & 0.96 $\pm$ 0.60 \\
C(H$\beta$)			& 0.1    & 0.095 $\pm$ 0.029 	 & 0.096 $\pm$ 0.027 \\
ABS(H~I)				& 1.0    & 1.34 $\pm$ 1.40 	 & 1.19 $\pm$ 1.07 \\
\enddata
\label{table:Integrated}
\end{deluxetable}

Since the temperature and reddening are the only two parameters directly impacting both the helium and hydrogen $\chi^{2}$, it might be expected that they would exhibit marked differences upon moving from sequential to integrated analysis.  Each does as do their most strongly coupled parameters, the density and underlying hydrogen absorption.  Integration improves the accuracy of the temperature marginally, but with a significantly larger uncertainty due to the added temperature dependence of the hydrogen lines. The reddening parameter, C(H$\beta$) is essentially unchanged.
The slight improvement for the reddening results from its weaker sensitivity to perturbation due to its equal impact upon the helium and hydrogen lines.  
Since both the reddening and the hydrogen absorption affect the weaker hydrogen lines more strongly, they are tightly coupled such that a more accurate and precise reddening correspondingly benefits the hydrogen absorption.  The less accurate density with accompanying larger uncertainty stems from the previously discussed temperature-density degeneracy.  The more uncertain temperature leads to the increased density uncertainty; while the non-linear nature of their trade-off biases the returned density.  The benefit of the more accurate parameters propagates to the abundance determination, improving from 0.08024 $\pm$ 0.00268 to 0.07995 $\pm$ 0.00324 (99.7\% to 99.9\%).  The greater variance in the density and temperature, however, leads to an increased error bar.  Overall, the effect of combining the helium and hydrogen calculations, on well-behaved spectra, is weak (and consistent at a set of higher and lower model parameters, though the sequential algorithm sometimes gives the marginally better abundance).  The most notable feature is the similarity in the returned results, excepting a more pronounced effect of degeneracy in the density and temperature uncertainty.

In the above synthetic analysis, both the sequential and integrated approaches achieve a well minimized hydrogen and helium $\chi^{2}$.  However, in real data, because the reddening and underlying hydrogen absorption both increase the flux ratio for the wavelengths shorter than H$\beta$ (and both decrease $\frac{H\alpha}{H\beta}$), minimization on the hydrogen lines alone may produce a relatively large $\chi^{2}$.  Then upon integration, the combination of the temperature dependent hydrogen emissivities, which were previously evaluated at the fixed [O~III] temperature, and the density-temperature degeneracy may lead the derived temperature to deviate from the sequential value.  The hydrogen $\chi^{2}$ exerts new pressure upon the temperature, but the weak (logarithmic) temperature dependence requires a large temperature adjustment.  As a result, the temperature can slide substantially to reduce the hydrogen $\chi^{2}$ while maintaining the helium $\chi^{2}$ by inversely adjusting the density.  The possibility for exactly this behavior is a primary example of the necessity of accounting for systematic error via Monte Carlo in calculating the abundance error.

\subsection{Integrating the Dataset}

\begin{figure}\centering  
\resizebox{\textwidth}{!}{\includegraphics{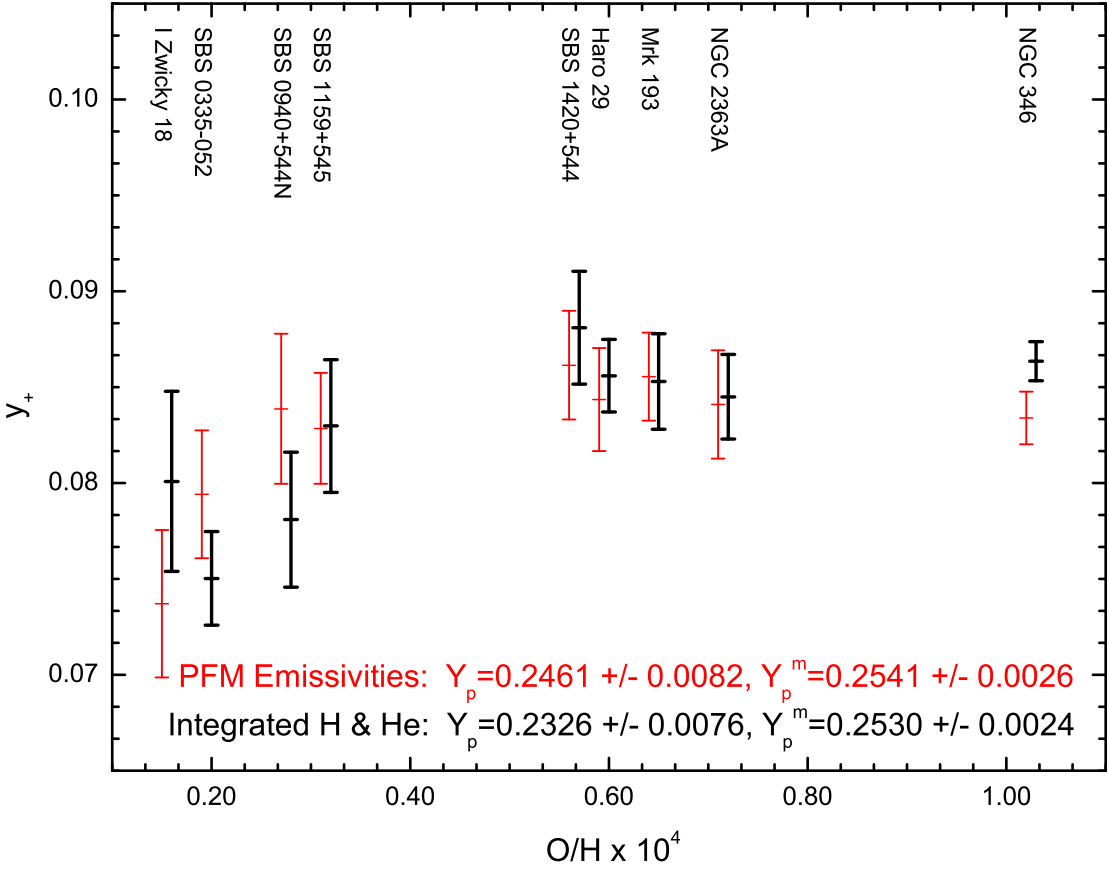}}
\caption{
Abundance comparison for the target objects in the sequential and integrated analyses.  The red thinner lines are for the sequential analysis (using PFM emissivities and shown in  Figure \ref{y_Progression_UE}) with the integrated values given in black thicker bars.  Again 
I~Zw~18 and NGC~346 are not used in calculating Y$_{p}$.
}
\label{y_Progression_Int}
\end{figure}

The resulting changes in the helium abundance in our nine objects due to 
the integration of the minimization procedure is shown in Figure \ref{y_Progression_Int}.
As evidenced in the preceding subsection, integration of the hydrogen and 
helium calculations could potentially yield significant changes to the 
temperature or reddening; however, for all objects, except one, the solved 
reddening was similar after integration to the result of \citetalias{os04}.  
The reddening for NGC~346 decreased from the \citetalias{os04} value of 0.174  
to 0.156, which is a difference of about two sigma.  This raises the helium 
abundance, as two of the three strong He lines are redder than H$\beta$.  

The temperature changed significantly post integration for the three most metal 
poor objects (see Figure \ref{y_Progression_Int}). I~Zw~18 showed a temperature 
increase (10,700 to 17,300 K) 
and a resulting increased abundance.  This difference in temperature is 
significant, and results in an increase in the helium abundance of about
1.5 sigma.  Note, however, that I~Zw~18 was not included in the original
``high quality'' dataset of Paper II because it did not meet two of the
criteria (EW(H$\beta$) and redshift), so this is likely due more to the
quality of the spectrum and the looser constraint on the derived temperature.
Nonetheless, the newly derived temperature is much more in line with that expected 
from the temperature derived from observations of the [O~III] emission lines.

However, the derived temperatures for SBS~0335-052 and SBS~940+544N both decreased 
significantly (16,200 to 10,700 K and 18,500 to 12,800 K respectively) 
and showed corresponding helium abundance decreases.  
The newly derived temperatures are significantly different from those
indicated by the [O~III] emission lines.  The magnitude of the effect 
on the abundance due to the temperature change for these three objects is 
apparent in Figure \ref{y_Progression_Int}.
As discussed above, these shifting temperatures result from the combination 
of the temperature dependent hydrogen emissivities with the density-temperature 
degeneracy.  Temperature changes allow the hydrogen $\chi^{2}$ another degree 
of freedom for minimization, and the density compensates within the helium $\chi^{2}$.  
This is supported by the fact that each of the above objects shows a larger 
contribution to the total $\chi^{2}$ from the hydrogen lines than from the
helium lines. 
Apart from inconsistency between the 
temperature and density used for the hydrogen based corrections, integration 
would necessarily decrease the total $\chi^{2}$, in validation of the procedure.  
The total $\chi^{2}$ of eight of nine objects does decrease (SBS 1159+545 
increases from 1.35 to 1.70), but the large parameter variation witnessed 
is of concern.  This provides further evidence for the importance of
addressing the temperature-density degeneracy.

\section{Underlying Absorption and Wavelength Dependence} \label{UAWD}

Stellar absorption underlying the helium emission lines has long been recognized
as potentially leading to underestimates of their measured fluxes. 
\citet{ds86, it98a, stt98} and \citet{os01} all demonstrated examples of how failure to 
correct for underlying stellar absorption in the He~I lines resulted in
underestimates of nebular helium abundances.
However, correcting for the effects of underlying He~I absorption is not
straightforward \citep[see discussions in][]{os01, its07}.
In \citet{os01}, the simplest possible approach was suggested: namely, the assumption
that the equivalent widths of all of the He~I absorption lines were identical
and could be obtained through the $\chi^2$ minimization for that value.  
The assumption of identical equivalent widths has merit since these lines are
usually saturated in the stars that are producing them, and so the main
problem is to essentially find a scaling factor to determine what fraction of the
continuum is produced by the stars with these saturated lines.
An alternative method is to use model stellar photospheres to produce
an estimate of the relative strengths of the underlying absorption.
In the second method, one needs estimates of the ages and metallicities
of the stars and, in addition, the fraction of the continuum due to the
stars.  Note that the large equivalent widths of the Balmer lines 
in the ``high quality'' sample indicate that a significant fraction of the 
continuum is nebular, so the assumption of a purely stellar continuum
is not entirely correct.

There is evidence for two significant departures from the simplest assumption
of identical equivalent widths of underlying He~I absorption.  First, observations
of stars indicate variations of order a factor of two \citep[see discussion in][]{os01}
between different He~I lines,
and this is also seen in the model stellar atmospheres.  Second, different 
populations of stars contribute differently to different parts of the spectrum
(i.e., young blue stars account for more of the blue continuum and older red
stars account for more of the red continuum).  If the stellar populations in
the galaxies observed were single epoch bursts, then one could neglect
the contributions to the continuum from different aged stellar populations.
However, it is becoming clear that the bursts of star formation that 
give rise to the bright H~II regions typically studied have durations of
order 100 Myr or longer \citep[][and references therein]{ms09}.
Since the He~I absorption lines are formed in the bluer stars, the red
continuum will have larger contributions from older populations, and thus,
the equivalent widths of the redder He~I lines will be weaker than 
predicted from models.

In order to investigate the possible limitations of assuming identical
equivalent widths of underlying stellar He~I absorption, we constructed models
of the emission from simple stellar populations using the Violent 
Star Formation Legacy tool ``SED@''\footnote{SED@ can be found at:
http://www.iaa.es/~mcs/sed@}.  The code used the input models from \citet{gcmlh05}
and \citet{mglc05}, and He~I absorption line strengths as a function of 
stellar population age were examined.  We then derived relative 
equivalent widths which were suitable over a large range of ages and these 
are presented here.  We also did this for the H~I absorption lines, although
the variation is much smaller.

The wavelength dependence of underlying absorption in terms of equivalent width 
for the hydrogen lines is normalized to H$\beta$ while He $\lambda$4471 is used 
for the helium lines.  Those normalization values then represent the varying 
physical model parameters, a$_{H}$ and a$_{He}$,
\begin{eqnarray}
a(H\alpha) = 0.942 ~~ a(H\beta) = 1.000 ~~ a(H\gamma) = 0.959 ~~ a(H\delta) =  0.896 \nonumber \\
a(3889) = 1.400 ~~~ a(4026) = 1.347 ~~~ a(4471) = 1.000 \nonumber \\
a(5876) = 0.874 ~~~ a(6678) = 0.525 ~~~ a(7065) = 0.400 .
\end{eqnarray}
The value at H9 is also needed to successfully deblend He($\lambda$3889).  
A linear fit to the four hydrogen values above provides a value of 0.882.

Our main goal here is to determine the relative importance of a secondary
correction to the assumption of identical equivalent widths for underlying
absorption.  Thus, the precise choice of the wavelength dependence of the
underlying absorption is not nearly as important as an assessment of the
effect.

\begin{figure}\centering  
\resizebox{\textwidth}{!}{\includegraphics{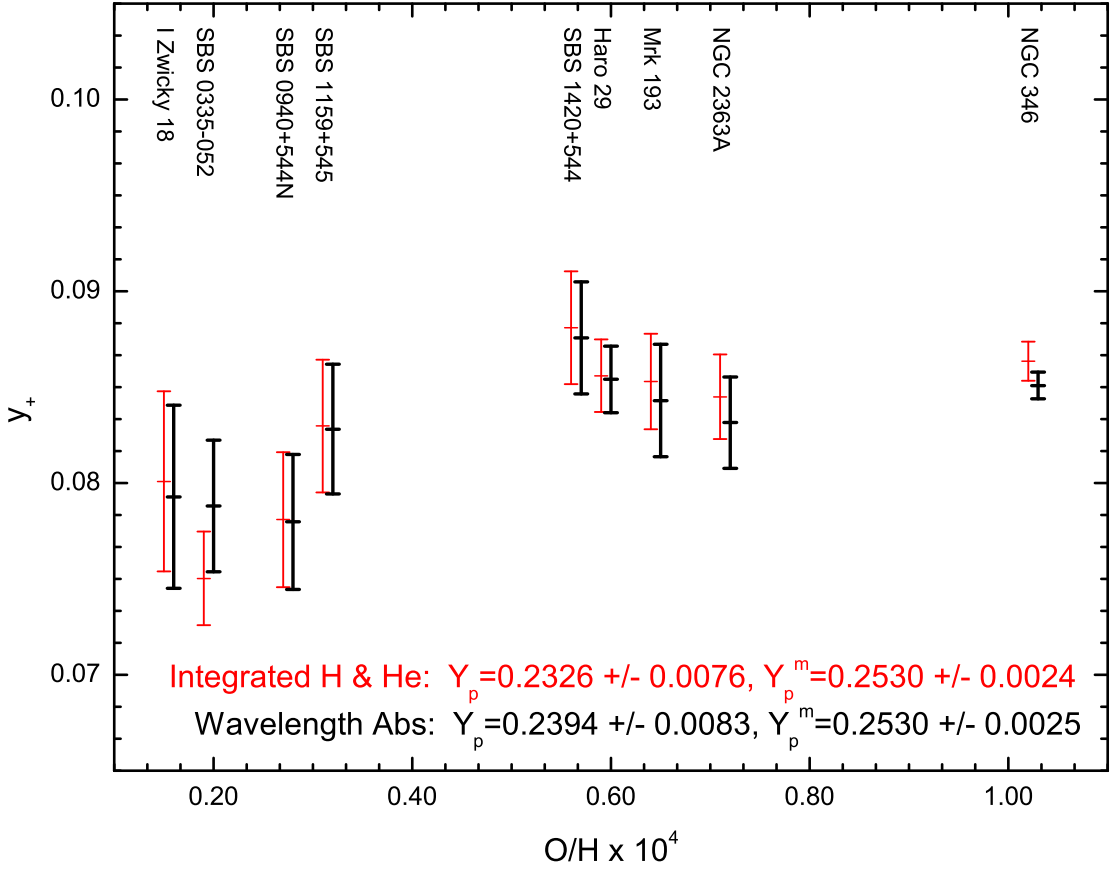}}
\caption{
Abundance comparison for the target objects upon inclusion of wavelength dependent absorption.  The red thinner lines are for the integrated analysis (of Figure \ref{y_Progression_Int}) with the same analysis performed including wavelength dependent absorption given in black thicker bars.  As before, I~Zw~18 and NGC~346 are not used in calculating Y$_{p}$.
}
\label{y_Progression_UA}
\end{figure}

Figure \ref{y_Progression_UA} shows the impact of assuming the
wavelength dependent absorption given above.  When comparing individual objects, 
in most cases there is no significant effect on the abundance. For most of
the objects there is a very slight downward shift, presumably due to smaller
corrections to the $\lambda$5876 and $\lambda$6678 lines.  Since this effect 
is only a small correction to the underlying absorption which itself is relatively 
weak, the minor impact is in agreement with expectations. Note that 
the impact is minor in large part due to the selection of targets for
the ``high quality'' sample.  By requiring large emission line equivalent
widths for the targets, we ensure that this intrinsically uncertain 
correction will be small.  

Note that there are two objects where the difference caused by the 
new assumption is noticeable: SBS~0335-052 and NGC~346.  In the case of
SBS~0335-052, we note that the solution changes with almost every 
small change in assumptions that we test.  This spectrum indicates a
significant amount of self absorption in the helium lines, and, for large
values of self-absorption our model is very uncertain.  For NGC~346, 
the uncertainties are very small, so even small changes in assumptions have 
noticeable effects on the solution.

\section{Neutral Hydrogen Collisional Emission} \label{NHCC}

To correct the measured hydrogen fluxes for emission due to neutral hydrogen collisional excitation, the relative amount of collisional to recombination emission for a given line can be calculated as follows,
\beq
\frac{C}{R}(\lambda) = \frac{n(H~I)(\sum_{i}K_{1 \rightarrow i}BR_{i \rightarrow j})BR_{j \rightarrow 2} n_{e}}{n(H~II) \alpha_{+ \rightarrow j}BR_{j \rightarrow 2} n_{e}} = \frac{n(H~I)}{n(H~II)}\frac{K_{eff}}{\alpha_{eff}} = {\xi}\frac{K_{eff}}{\alpha_{eff}},
\eeq
where \textit{K} represents the collisional transition rate from the ground state to some level, \textit{i}, above the transition level of interest, \textit{j} ($j \rightarrow 2 = \lambda$).  Downward transitions from \textit{i} occur with a variety of branching paths so \textit{$BR_{i \rightarrow j}$} represents the relevant branching fraction for transitioning ultimately to \textit{j}.  \textit{$\alpha$} is the effective recombination rate to level \textit{j} (\textit{$\alpha_{+ \rightarrow j}=\alpha_{eff}$} with ``$+ \rightarrow j$'' serving as an analogue to \textit{$K_{1 \rightarrow i}$} but for ionization recombining to level \textit{j}).  \textit{$K_{eff}$} is the sum over the excitation levels of the collisional transition rate, \textit{$K_{1 \rightarrow i}$}, times the corresponding branching fraction, \textit{$BR_{i \rightarrow j}$}.  Given the relative complexity of the definition of \textit{$\frac{C}{R}$}, it is useful to note that the electron density and branching fraction from level \textit{j} to 2 cancel, and we are left with the ratio of the neutral hydrogen density, n(H~I), to the ionized hydrogen density, n(H~II), times a collisional rate over a recombination rate.  The ratio $\frac{n(H~I)}{n(H~II)}$ is then defined as the model parameter $\xi$ which will be solved via the minimization.  

As defined above, \textit{$\frac{C}{R}$} makes the simplifying but very accurate assumption, at these temperatures, that all of the neutral hydrogen is excited from the ground state.  The collisional rates are based on power law fits ($A T^{B}$) to the effective collisional strengths, $\Upsilon$, reported in \citet{and02} over the temperature range 5000 to 35,000 K.  The effective collisional strength as defined below allows for more convenient fitting due to its more gradual temperature dependence and is easily converted to the collisional excitation rate via the accompanying rate formula.
\beq
\Upsilon_{ij}=\int_{0}^{\infty}\Omega(i \rightarrow j)\exp(-\frac{\epsilon_{j}}{k_{B}T})d(\frac{\epsilon_{j}}{k_{B}T})
\eeq
\begin{eqnarray}
K_{i \rightarrow j}=\frac{2\sqrt{\pi}\alpha ca_{0}^{2}}{\omega_{i}}\sqrt{\frac{I_{H}}{k_{B}T}}\exp(-\frac{\triangle E_{ij}}{k_{b}T})\Upsilon_{ij} \\
K_{1 \rightarrow i}=4.004\times10^{-8}\sqrt{\frac{1}{k_{B}T}}\exp(\frac{-13.6(1-\frac{1}{i^{2}})}{k_{b}T})\Upsilon_{1i} \nonumber
\end{eqnarray}
with $\Omega$, the collisional strength, related to the cross-section by $\Omega_{ij}=\frac{2m\epsilon_i\omega_{i}}{\pi\hbar^{2}}\sigma_{ij}(\epsilon_i)$, $\epsilon_j$ is the final outgoing electron energy, $\epsilon_i$ is the incident electron energy, $I_{H}=-13.6~eV$ the ionization potential, $a_0$ is the Bohr radius, and $\omega_{i}=(2s+1)(2l+1)$ the statistical weight of the level.
The branching fractions are calculated directly from the Einstein transition coefficients, and the effective recombination rate is fit from the case B data of Hummer \& Story (1987) for an electron density of 100 per cm$^{3}$ and over the temperature range 10,000 to 30,000 K.  Again a power law fit is chosen for simplicity and models well.  In theory, the collisional sum includes an infinite number of levels, but the probabilities fall off rather quickly.  For this work the sum excludes any terms contributing less than 1\%, giving the following contributions:\\
\indent For H$\alpha$, 3s, 3p, 3d, 4s, 4d, and 4f are included;\\
\indent For H$\beta$, 4s, 4p, 4d, 4f, 5s, 5d, 5f, and 5g;\\
\indent For H$\gamma$, 5s, 5p, 5d, 5f, and 5g;\\
\citet{and02} only report values for collisions up to a principle quantum number of 5; therefore, to calculate the contribution to H$\delta$, the equation for H$\gamma$ was used and scaled by the appropriate energy level difference,
\beq
\frac{C}{R}(\delta) = \frac{C}{R}(\gamma)\exp(\frac{-13.6\,eV (\frac{1}{5^{2}}-\frac{1}{6^{2}})}{k_{B}T}).
\eeq

Table \ref{table:HIcoll} lists the coefficients for constructing $\frac{K_{eff}}{\alpha_{eff}}$, the quantity that multiplies $\xi$ and provides the increasing collisional enhancement factor for increasing temperature ($T_{4} = \frac{T}{10^{4}}$).
\beq
\frac{K_{eff}}{\alpha_{eff}} = \sum_{i}a_{i}\exp(-\frac{b_{i}}{T_{4}}){T_{4}}^{c_{i}}
\eeq
As can be clearly seen in Figure \ref{H_CR_T} below, the collisional excitation grows strongly with temperature.  This has the effect of leaving $\xi$ poorly constrained at 
low temperature, but potentially well constrained at high temperature due to its 
now significant effect on the hydrogen $\chi^{2}$.  The neutral hydrogen collisional 
correction has no direct effect on the helium $\chi^{2}$; however, it will directly 
raise the abundance due to the decrease in the normalizing H$\beta$ flux.
In this regard, when working with lower temperature objects (like NGC~346) the
program is vulnerable to producing non-physical results (very high neutral
fractions).  Here we will analyze NGC~346 for demonstration purposes, but
we do not use NGC~346 in our final determination of the primordial helium abundance.

\begin{deluxetable}{lccccccccc}
\tabletypesize{\footnotesize}
\tablecaption{Coefficients for the neutral hydrogen collisional correction}
\tablewidth{0pt}
\tablehead
{
  \colhead{Line}                    & 
  \colhead{Coefficient}             &
  \colhead{Terms}                   \\
}
\startdata
H$\alpha$ & $a$ & 0.4155 & 2.4965 & 2.4063 & 0.2914 & 0.3685 & 4.6426 \\
          & $b$ & -14.80 & -14.03 & -14.03 & -14.80 & -14.80 & -14.03 \\
          & $c$ & 0.4209 & 0.5853 & 0.6187 & 0.6766 & 0.7076 & 0.7788 \\
\hline
H$\beta$ & $a$ & 0.2384 & 0.6964 & 0.1991 & 0.1409 & 0.2201 & 1.9228 & 1.4845 & 2.8179 \\
          & $b$ & -15.15 & -14.80 & -15.15 & -15.15 & -15.15 & -14.80 & -14.80 & -14.80 \\
          & $c$ & 0.3082 & 0.4978 & 0.6017 & 0.6765 & 0.7293 & 0.7535 & 0.7845 & 0.9352 \\
\hline
H$\gamma$ & $a$ & 0.3629 & 0.8351 & 2.0044 & 1.4757 & 2.7947 \\
          & $b$ & -15.15 & -15.15 & -15.15 & -15.15 & -15.15 \\
          & $c$ & 0.3598 & 0.6533 & 0.7281 & 0.7809 & 0.8582 \\
\hline
H$\delta$ & $a$ & 0.3629 & 0.8351 & 2.0044 & 1.4757 & 2.7947 \\
          & $b$ & -15.34 & -15.34 & -15.34 & -15.34 & -15.34 \\
          & $c$ & 0.3598 & 0.6533 & 0.7281 & 0.7809 & 0.8582 \\
\enddata
\label{table:HIcoll}
\end{deluxetable}

\begin{figure}
\resizebox{\textwidth}{!}{\includegraphics{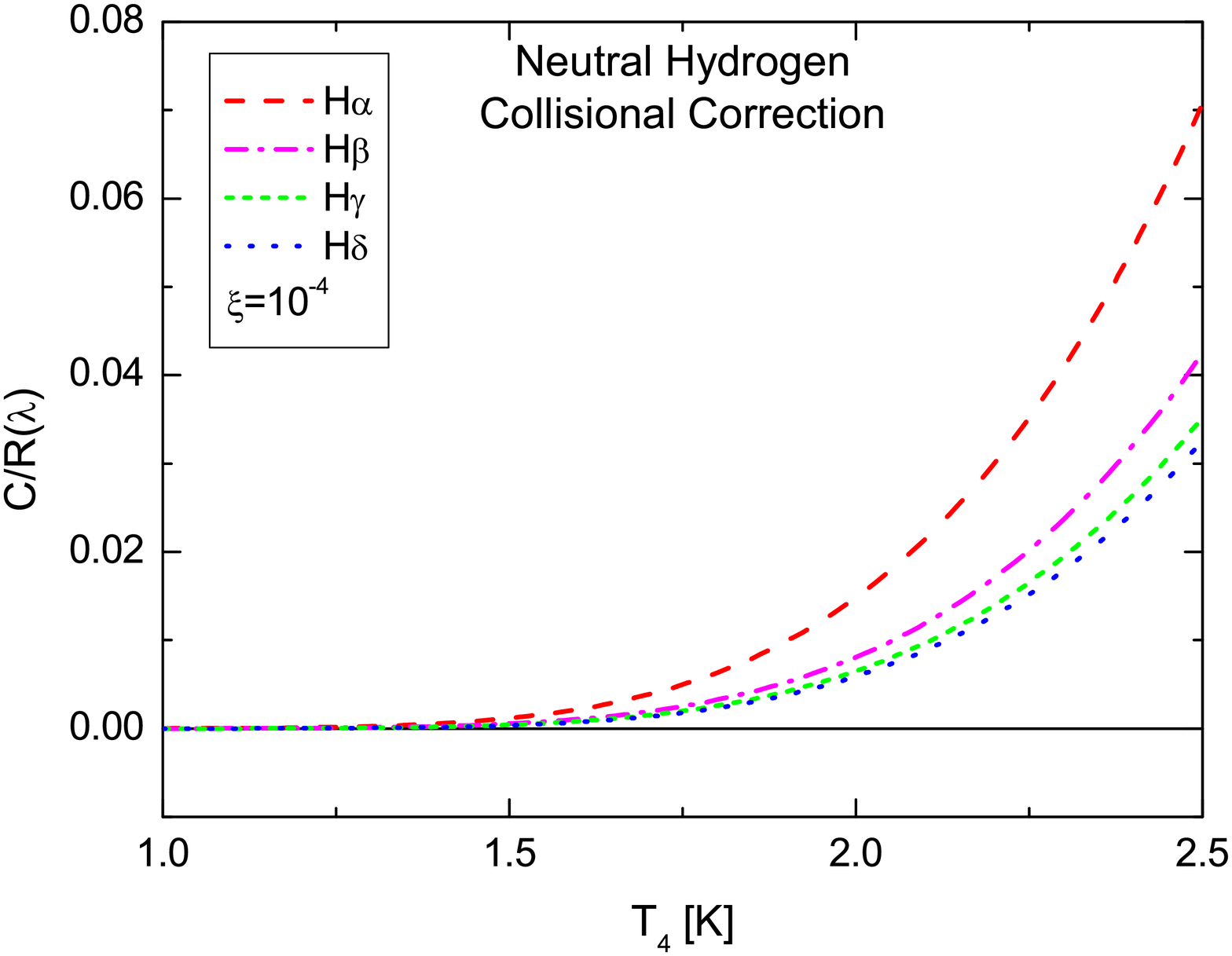}}
\caption{
The collisional correction factor for four hydrogen emission lines due to neutral hydrogen.  The behavior is dominantly exponential with temperature, with, as expected, larger corrections for the lower energy emission lines.  Note that the enhancement will scale linearly with increasing the neutral to ionized hydrogen ratio, $\xi$ (a characteristic value of 10$^{-4}$ is used for the plot).
}
\label{H_CR_T}
\end{figure}

\subsection{A Return to Integration}

The effects of integrating the helium and hydrogen $\chi^{2}$ minimizations were already discussed in section \ref{Int_Investigation}.  However, apart from the general motivation of increased consistency, the inclusion of the neutral hydrogen fraction provides a strong practical benefit.  Again, synthetic testing, with its known parameters, provides insight.  The same generating parameters as in section \ref{Int_Investigation} are used with the addition of $\xi$ = 1.0 x 10$^{-4}$.  The generating and solved parameters and abundance are listed in Table \ref{table:Integrated_NHCC}.

\begin{deluxetable}{lccc}
\tabletypesize{\footnotesize}
\tablewidth{0pt}
\tablecaption{Comparing Sequential and Integrated Analyses with Synthetic Data}
\tablehead{
&
\colhead{Input}		 &
\colhead{Sequential}     &
\colhead{Integrated}}
\startdata
He$^+$/H$^+$			& 0.08 	 & 0.08375 $\pm$ 0.00502 & 0.08099 $\pm$ 0.00373 \\
T$_e$				& 18,000 & 17,138 $\pm$ 1812 	& 17,505 $\pm$ 3103 \\
N$_e$				& 100.0  & 141.9 $\pm$ 252.7 	& 173.7 $\pm$ 311.4 \\
ABS(He~I)			& 1.0	 & 1.16 $\pm$ 0.10 	& 1.05 $\pm$ 0.10 \\
$\tau$				& 1.0    & 1.35 $\pm$ 0.42 	& 1.15 $\pm$ 0.68 \\
C(H$\beta$)			& 0.1    & 0.05 $\pm$ 0.05 	& 0.08 $\pm$ 0.03 \\
ABS(H~I)				& 1.0    & 2.16 $\pm$ 1.65 	& 1.32 $\pm$ 1.10 \\
$\xi$ $\times$ 10$^4$   	& 1.0    & 11.62 $\pm$ 12.22 	& 60.65 $\pm$ 269.67 \\
\enddata
\label{table:Integrated_NHCC}
\end{deluxetable}

As before, the most significant changes are seen in the temperature, density, reddening, and underlying hydrogen absorption.  Now, the addition of neutral hydrogen requires solving three parameters (C(H$\beta$), a$_{H}$, $\xi$) from the hydrogen lines alone if sequential analysis is used, but only two for integrated analysis (since the reddening affects both the hydrogen and helium lines).  This dramatically improves the reddening accuracy, 52\% to 82\%.  The aforementioned similarity of wavelength behavior between the reddening and hydrogen absorption again leads to corresponding improvement for the hydrogen absorption.  The much larger and more discrepant neutral hydrogen fraction, with an especially increased uncertainty, is, on the surface, disconcerting.  The increased temperature uncertainty is the cause.  As discussed in the previous section, at low temperatures the neutral hydrogen collisional emission rate is greatly reduced permitting a larger fraction of neutral hydrogen.  Therefore the introduction of lower temperature solutions after integration leads to lower neutral hydrogen collisional emission rate which is then compensated by a higher fraction of neutral hydrogen.  The resulting correction is actually smaller for the integrated analysis (e.g., $\xi \frac{C}{R}(H\beta)=3.3\% \rightarrow 1.4\%$) such that the neutral hydrogen correction is improved from the sequential to integrated analysis.  The full benefit of the more accurate solution is manifested in the abundance, improving from 0.08375 $\pm$ 0.00502 to 0.08099 $\pm$ 0.00373 (95\% to 99\%).  

To summarize, the use of the solved temperature in evaluating the hydrogen lines, though necessary for consistency, exposes the neutral hydrogen fraction to the temperature-density degeneracy, as evidenced by the uncertainty on the neutral hydrogen fraction.  Apart from this temperature feedback, however, integration clearly benefits the neutral hydrogen solution as well as the other parameters.  Primarily, the reddening, constrained through the helium and hydrogen lines, is more precisely determined; therefore, the underlying absorption and neutral hydrogen fraction are allowed less range in correcting the hydrogen lines, and the abundance determination benefits.

\subsection{The Dataset Analysis with the New Neutral Hydrogen Parameter}

\begin{figure}
\resizebox{\textwidth}{!}{\includegraphics{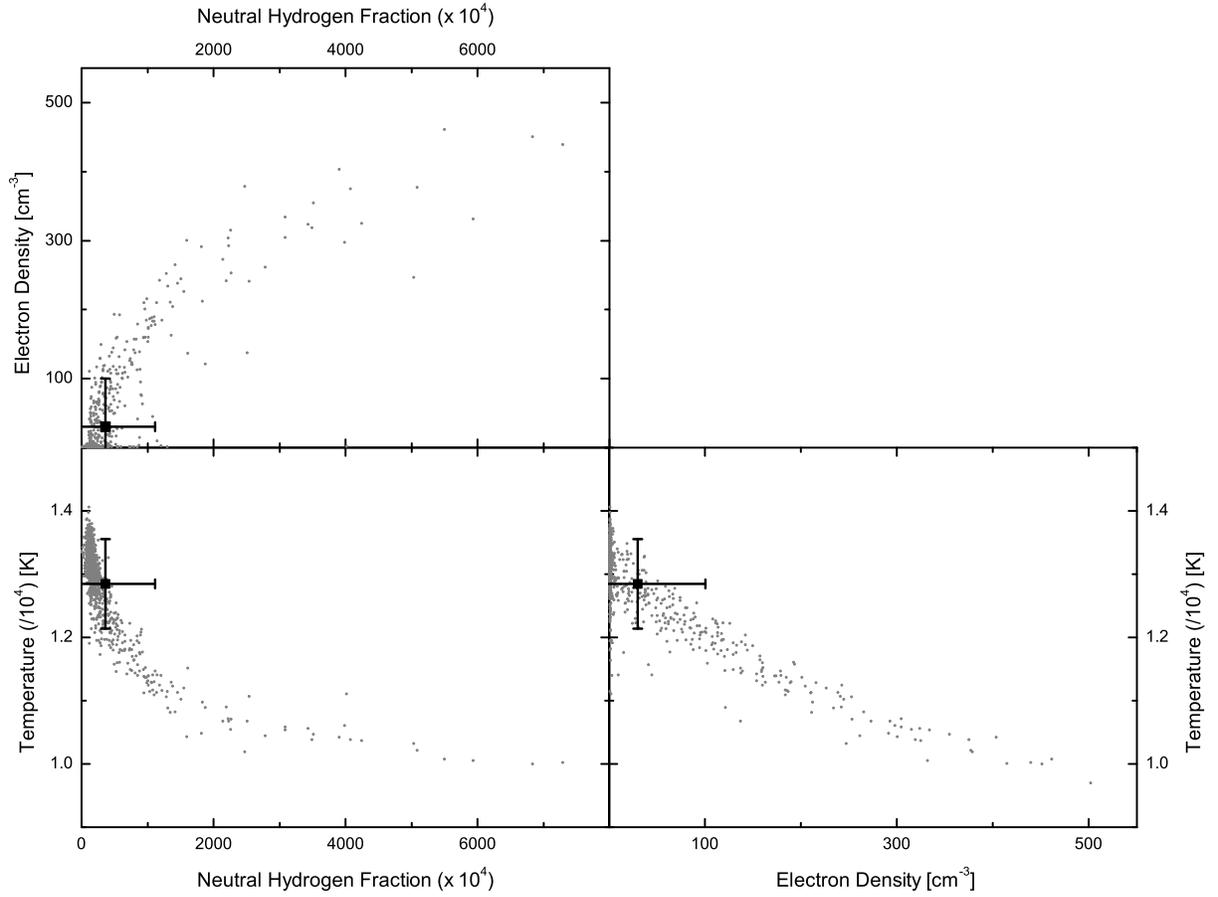}}
\caption{
Monte Carlo plots (1000 points) of density, temperature, and neutral hydrogen fraction for NGC~346.  The strong correlation between the three parameters demonstrates the difficulty in solving for them independently.
}
\label{346DTHI}
\end{figure}

The introduction of the neutral hydrogen collisional emission complicates the abundance determination.  A three way degeneracy between electron density, temperature, and the neutral hydrogen fraction emerges.  These degeneracies are shown in Figure \ref{346DTHI} which shows the
Monte Carlo results for NGC~346.  Plotted are the individual results for $n_e, T$, and $\xi$ over the 1000 MC realizations. As one can see, the solutions map out a wide range for these parameter values. 

\begin{figure}\centering  
\resizebox{\textwidth}{!}{\includegraphics{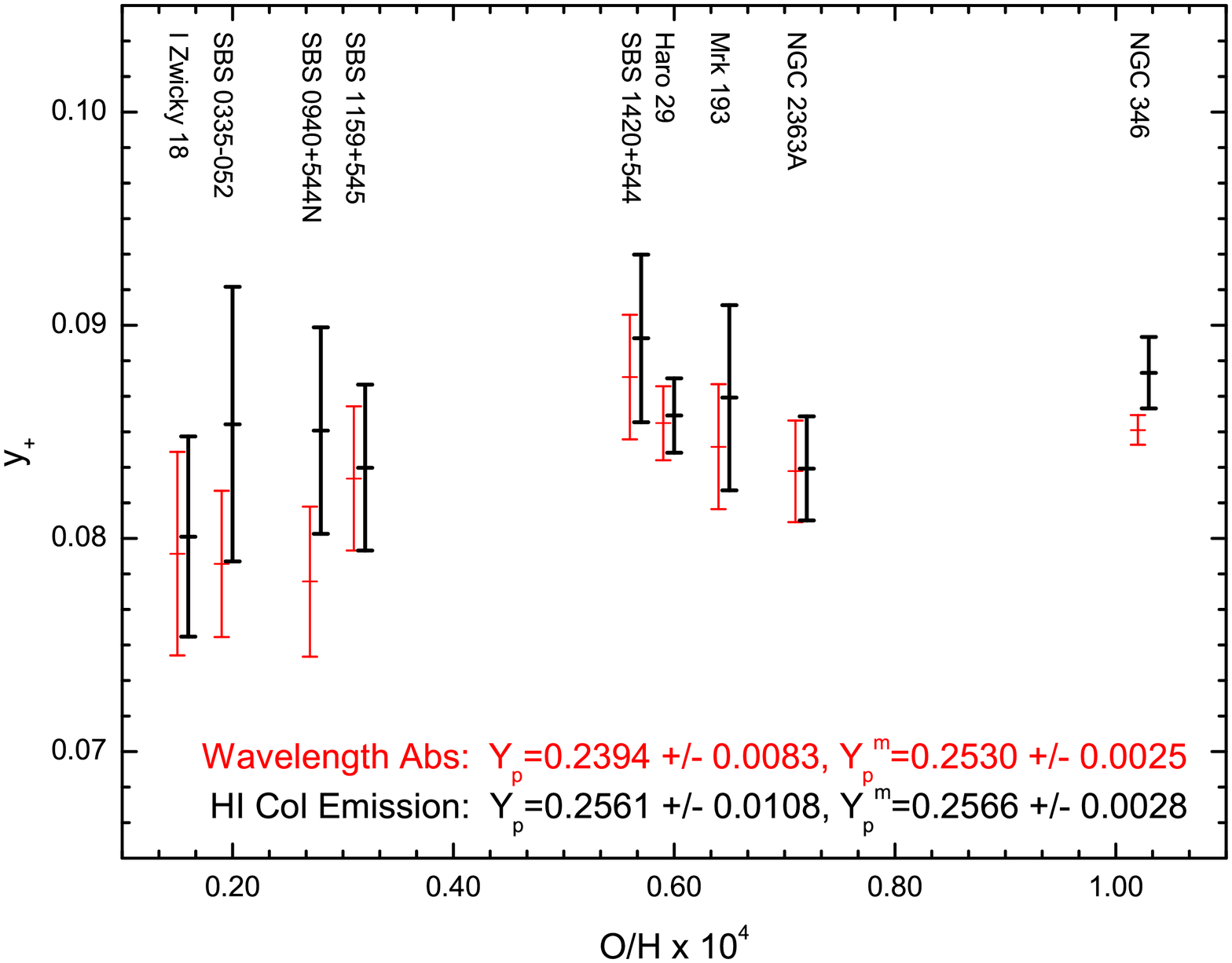}}
\caption{
Abundance comparison for the target objects detailing the effect of the neutral hydrogen collisional emission.  The red thinner lines are for the integrated analysis including wavelength dependent absorption (of Figure \ref{y_Progression_UA}) with the same analysis performed including neutral hydrogen collisional emission given in black thicker bars.  As before, 
I~Zw~18 and NGC~346 are not used in calculating Y$_{p}$.  }
\label{y_Progression_NH}
\end{figure}

Since the neutral hydrogen fraction multiplies  \textit{$\frac{K}{\alpha}$} directly, the neutral hydrogen fraction can be increased to compensate for a decrease in the collisional to recombination rate which is itself only temperature dependent.  Furthermore the neutral hydrogen collisional emission correction can have a surprisingly significant effect in raising the abundance as evidenced in Figure \ref{y_Progression_NH}.  Since the collisional correction for H$\beta$ multiplies each helium line abundance, it multiplicatively raises the abundance but does not change the value of the corresponding $\chi^{2}$.  As a result, the abundance of all nine objects increases as does the derived primordial helium mass fraction.  

\subsection{Comparing Implementations}

The approach detailed above differs from the recent analogous additions 
to ITS07 and PLP07 in several important ways.  ITS07 varies the collisional 
contribution to H$\alpha$ between 0 and 5\% and takes the collisional 
contribution to H$\beta$, and therefore to y$^{+}$, to be $\frac{1}{3}$ as large.  
The minimum of the reddening corrected helium $\chi^{2}$ (each minimized 
over n$_{e}$, $\tau$, and T) is taken to be the solution.  This method does 
not allow for a full sampling of possible parameter values as 
we find in the MC analysis exhibited in Figure \ref{346DTHI}.
Regarding the neutral hydrogen collisional emission, the key differences 
with this work are modeling the relative collisional contributions to 
H$\alpha$ and H$\beta$ only in determining the reddening and treating the 
collisional emission fraction in its entirety as a model parameter rather 
than a neutral hydrogen fraction times a temperature dependent relative 
collisional emission rate.  

PLP07 creates a photoionization model for each of their objects using 
CLOUDY which outputs the collisional contributions to the hydrogen lines.  
Their model also makes use of the collisional strength data of Anderson \etal (2002).  
The reddening is calculated from the corrected H$\alpha$ and H$\beta$, 
and y$^{+}$ is increased by the decrease in H$\beta$.  The primary differences 
with the neutral hydrogen emission of this work are, again, correcting only 
H$\alpha$ and H$\beta$ in determining the reddening and the construction of a 
photoionization model, in particular requiring the specification of ionizing 
radiation and geometry of the region, to determine the neutral hydrogen 
collisional contribution.  These differences extend beyond the neutral 
hydrogen collisional contribution modeling and determination. We emphasize that, 
here, in addition to the collisional contributions to H$\gamma$ and H$\delta$,
all six He lines, due to the combined hydrogen and helium calculations, 
are included in determining the reddening as well.
Furthermore, the integrated approach utilizes the same temperature in determining 
the hydrogen recombination emission, hydrogen collisional emission, and the 
helium emission.

Table \ref{table:CRHI} below details the relative collisional to recombination emission contributions, \textit{$\frac{C}{R}$}, for the objects here analyzed with corresponding corrections of ITS07 and PLP07 listed where available.  Note that for this work \textit{$\frac{C}{R}$} is computed for each Monte Carlo realization from the solved neutral hydrogen fraction, $\xi$, and temperature, T, with the table listing the Monte Carlo average.  The magnitude of the corrections calculated for this work is in broad agreement with that found in ITS07 (for which the solved H$\alpha$ correction is restricted to less than 0.05) and PLP07.  The results for individual objects, however, do not in general agree.  The most notable disagreements are for Haro~29 and NGC~2363A, where this work finds a much smaller correction than either ITS07 or PLP07, and for Mrk~193, much larger than ITS07.  Close agreement is found for H$\alpha$ for SBS~0335-052 with PLP07 and SBS~0940+544N, SBS~1159+545, and SBS~1420+544 (only reported by ITS07).  As evidenced by those last three objects, the temperature dependent neutral hydrogen collisional contribution of H$\beta$ relative to H$\alpha$ is always less than $\frac{1}{3}$, the value used by ITS07.  PLP07 finds this ratio of the H$\beta$ to H$\alpha$ contribution always greater than this work but also always less than ITS07.  Both this work and PLP07 calculate relative corrections to the H$\alpha$ emission in excess of the 5\% limit imposed by ITS07 for several objects.  Overall, the different neutral hydrogen emission models yield qualitatively and quantitatively different results, but the variance and magnitude of the effect are in accordance.

\begin{landscape}
\begin{deluxetable}{lcccccccc}
\tabletypesize{\scriptsize}
\tablewidth{0pt}
\tablecaption{Comparison of Neutral Hydrogen Collisional Contribution}
\tablehead{
\colhead{$\frac{C}{R}(\lambda)$:} &
\colhead{H$\alpha$}     &
\colhead{H$\beta$}     &
\colhead{H$\gamma$}     &
\colhead{H$\delta$}     &
\colhead{H$\alpha$}     &
\colhead{H$\beta$}     &
\colhead{H$\alpha$}     &
\colhead{H$\beta$} \\
\hline
\colhead{Object} &
\multicolumn{4}{c}{This Analysis} &
\multicolumn{2}{c}{ITS07} &
\multicolumn{2}{c}{PLP07\tablenotemark{a}}}
\startdata
I~Zw~18 & 0.0070 $\pm$ 0.0089 & 0.0037 $\pm$ 0.0083 & 0.0030 $\pm$ 0.0083 & 0.0027 $\pm$ 0.0083 & 0.0186 & 0.0222 & 0.075 & 0.056 \\
SBS~0335-052 & 0.0881 $\pm$ 0.0607 & 0.0425 $\pm$ 0.0464 & 0.0334 $\pm$ 0.0488 & 0.0297 $\pm$ 0.0501 & 0.0033 & 0.0036 & 0.094 & 0.071 \\
SBS~0940+544N & 0.0545 $\pm$ 0.0323 & 0.0273 $\pm$ 0.0223 & 0.0217 $\pm$ 0.0237 & 0.0194 $\pm$ 0.0247 & 0.0499 & 0.0490 & - & - \\
SBS~1159+545 & 0.0542 $\pm$ 0.0342 & 0.0262 $\pm$ 0.0244 & 0.0205 $\pm$ 0.0259 & 0.0182 $\pm$ 0.0268 & 0.0492 & 0.0452 & - & - \\
SBS~1420+544 & 0.0579 $\pm$ 0.0373 & 0.0309 $\pm$ 0.0281 & 0.0249 $\pm$ 0.0293 & 0.0226 $\pm$ 0.0301 & 0.0497 & 0.0520 & - & - \\
Haro~29 & 0.0011 $\pm$ 0.0016 & 0.0006 $\pm$ 0.0015 & 0.0004 $\pm$ 0.0016 & 0.0004 $\pm$ 0.0016 & 0.0357 & 0.0426 & 0.034 & 0.021 \\
Mrk~193 & 0.0442 $\pm$ 0.0400 & 0.0191 $\pm$ 0.0339 & 0.0145 $\pm$ 0.0348 & 0.0126 $\pm$ 0.0353 & 0.0013 & 0.0004 & - & - \\
NGC~2363A & 0.0013 $\pm$ 0.0069 & 0.0005 $\pm$ 0.0069 & 0.0004 $\pm$ 0.0069 & 0.0004 $\pm$ 0.0069 & 0.0115 & 0.0156 & 0.038 & 0.028 \\
NGC~346 & 0.0461 $\pm$ 0.0282 & 0.0193 $\pm$ 0.0164 & 0.0145 $\pm$ 0.0181 & 0.0125 $\pm$ 0.0191 & - & - & 0.011 & 0.007 \\
\enddata
\tablenotetext{a}{In terms of the variable used in PLP07, $x_{\lambda}=\frac{I(\lambda)_{col}}{I(\lambda)_{tot}}$, $\frac{C}{R}(\lambda)=\frac{x_{\lambda}}{1-x_{\lambda}}$}
\label{table:CRHI}
\end{deluxetable}
\end{landscape}

\section{Ionization Correction Factors} \label{icfs}

If the ionized hydrogen is not spatially co-existent with the ionized helium, then
a correction needs to be made for this effect.  This can happen for two different
reasons.  If the ionizing radiation field lacks sufficient high energy photons, 
then some of the helium will be neutral where the hydrogen is ionized.  It is
also possible that lower energy ionizing photons will be preferentially
captured by hydrogen atoms resulting in regions where helium is ionized and 
hydrogen is not (this can occur at the edges of 1-dimensional photoionization
models of H~II regions).  Since neutral inclusions are impossible to 
detect in spectra of unresolved H~II regions,  corrections for these
effects are inherently uncertain.  Typically photoionization models are 
used as a guideline for the importance of this correction.

The consensus approach to dealing with neutral helium has been to 
observe objects with spectra that indicate that the presence of
neutral helium is highly unlikely.  
Based on photoionization models,
\citet{pag92} established a criterion to select only H~II regions
with sufficiently hard radiation fields, and this method has been
followed through most investigations.

The reverse correction, dealing with the presence of neutral hydrogen, has
been discussed \citep{vgs00, bfm00, gsv02}, but generally is not applied nor
accounted for in the uncertainty.  Interestingly, \citet{bfm00} found that 
by eliminating H~II regions with relatively strong [O~I] $\lambda$6300
emission, that the scatter in helium abundances decreased, and interpreted
this as a sign of neutral hydrogen within the ionized helium.
It is also possible that the [O~I] $\lambda$6300 emission is indicative
of collisional excitation due to the presence of supernova remnants
\citep[e.g.,][]{s85}, and that the elevated $\lambda$6300 emission could
simply be an indicator that the nebula is not excited purely by
photoionization, and thus the simple model used to derive abundances
is not appropriate.

In this regard, detections  of collisionally excited Balmer emission 
are of interest.  If as much as 1\% of the helium is co-existent with
neutral hydrogen, then collisional excitation of the lower Balmer 
emission lines would be important in that volume.  To our knowledge,
none of the studies of these ionization correction factors ever
looked into detecting the neutral hydrogen by looking at the 
collisionally excited Balmer emission.  Although it is beyond the
scope of this work, we believe that this would be a worthwhile pursuit.
For the present, we estimate that the amplitude and the uncertainty 
in the ionization correction are much smaller than the other 
factors discussed in this paper.

\section{Primordial Helium} \label{Results}

Figures \ref{y_Progression_UE}, \ref{y_Progression_Int}, \ref{y_Progression_UA}, \ref{y_Progression_NH} summarize the results of each subsequent change or addition of sections \ref{Determining y}-\ref{NHCC}.  Table \ref{table:GTO} provides numerical results comparing \citetalias{os04} and the analysis of this work with all of the changes and additions included.  The abundance and all seven (except the neutral hydrogen fraction for \citetalias{os04}) parameters are listed.  I~Zw~18 was not analyzed in \citetalias{os04} and is therefore not listed.  The cumulative effect of the updated emissivities, integration, wavelength dependent absorption, and neutral hydrogen collisional emission is to raise the abundance for seven of the eight objects.  SBS~1159+545 undergoes the previously mentioned decrease from the updated emissivities but is then very stable.  The abundance error increases for six of the eight with Haro~29 and NGC~2363A stable after integration decreased their error.  Mutual objects with ITS07 and PLP07 allows for further comparison (see Table \ref{table:ITS_PLP}).  Broadly, this work finds lower temperatures, higher densities, and lower optical depths than ITS07.  There is no clear pattern with the abundance though on average it is higher.  Our derived densities are lower than those of PLP07 with correspondingly higher abundances.  The only object not in agreement is NGC~346.  The solution for NGC~346 exhibits susceptibility to large variance, relative to its small uncertainty.

\begin{figure}
\resizebox{\textwidth}{!}{\includegraphics{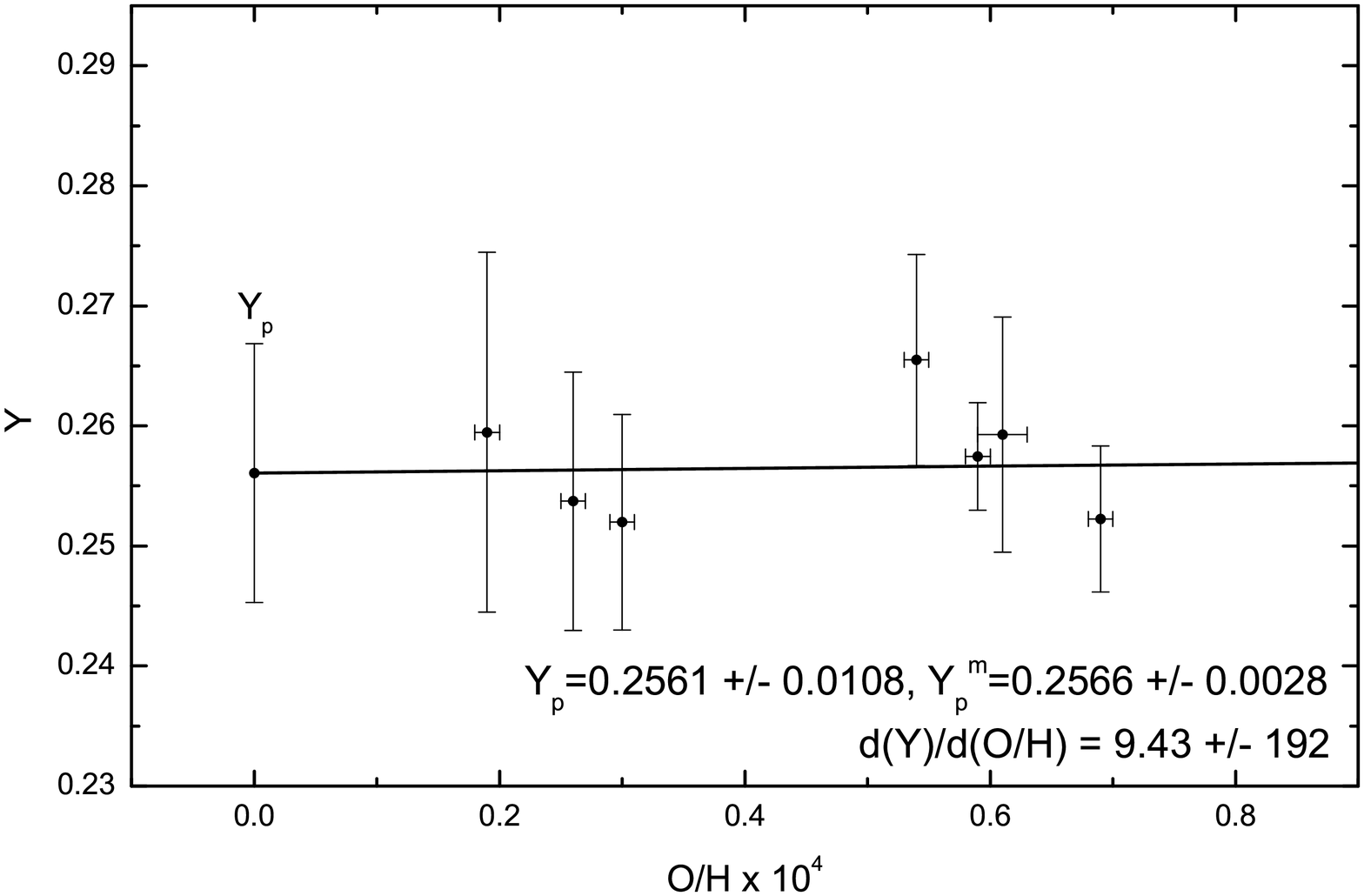}}
\caption{
Helium abundance (mass fraction) versus oxygen to hydrogen ratio regression calculating the primordial helium abundance.
}
\label{Y_OH}
\end{figure}

Our primary interest is the primordial helium abundance (mass fraction), Y$_{p}$.  To extrapolate back to a primordial value, the oxygen to hydrogen mass fraction, O/H, calculated in \citetalias{os04}, is used in conjunction with Y, the helium mass fraction, including a correction for  y$^{++}$.\footnote{This work takes $Z=20(O/H)$ such that $Y=\frac{4y(1-20(O/H))}{1+4y}$}  The relevant values are listed in Table \ref{table:PH}.  \citetalias{os04} analyzed NGC~346 but did not include it in the regression because its more evolved chemical nature invokes more error in assuming a linear relationship between y$^{+}$ and O/H.  I~Zw~18 was not analyzed in \citetalias{os04} both because of its low equivalent width for H$\beta$ ($EW(H\beta)=135<200$\AA) making it susceptible to large corrections for underlying absorption and because of it's radial velocity making it susceptible to contamination by Galactic Na~I absorption.  A linear regression of Y for the seven targets of \citetalias{os04} versus O/H yields,
\beq
Y_p = 0.2561 \pm 0.0108,
\label{eq:Yp}
\eeq
with slope 9 $\pm$ 192 and $\chi^{2}$ = 2.0.  This result is shown in Figure \ref{Y_OH}.  Given the large uncertainty, this agrees well with the WMAP result of $Y_p = 0.2486 \pm 0.0002$.  \citetalias{os04} found $Y_p = 0.2495 \pm 0.0092$.  The increased intercept and error of this work is directly attributable to y$^{+}$ increasing for six targets and an uncertainty increase for four.  The slope is similar to that
found in \citetalias{os04} (54 $\pm$ 187) and still consistent with zero.  Given the short baseline, a mean evaluation yields,
\beq
 Y_p = 0.2566 \pm 0.0028
 \eeq
 for seven which is very similar to the Bayesian result \citep{hog} of $0.2563 \pm 0.0028$.  

Because of the large uncertainty in Y,  Equation (\ref{eq:Yp}) is in agreement with PLP07, $Y_p = 0.2477 \pm 0.0029$, using PFM, and ITS07, $Y_p = 0.2516 \pm 0.0011$, using BSS rescaled to PFM emissivities.  The striking, larger error found here (as in Paper II) in comparison to PLP07 and ITS07 is attributable to use of an unrestricted Monte Carlo, the application of all systematic corrections simultaneously, and a smaller population of objects (than ITS07).   I~Zw~18 is below the regression line and, being the lowest metallicity point, it would lower the intercept and raise the slope. NGC~346
would extended the O/H domain and decrease the uncertainty in both the intercept and slope.  If both regions are included, we would find, $Y_p = 0.2528 \pm 0.0060$, with a slope of $64 \pm 80$.  

As noted earlier, the inclusion of the neutral hydrogen collisional correction has a large impact on the result. As seen in Figure \ref{y_Progression_NH}, the same seven object regression performed using the analysis with $\xi$ = 0 yields a lower tighter intercept of, $Y_p = 0.2394 \pm 0.0083$. Though not desirable, the spread of the results with and without the neutral hydrogen correction highlights the dominance of systematic effects and therefore the necessity of quantifying the full effect the associated error.

\begin{figure}\centering
\resizebox{\textwidth}{!}{\includegraphics{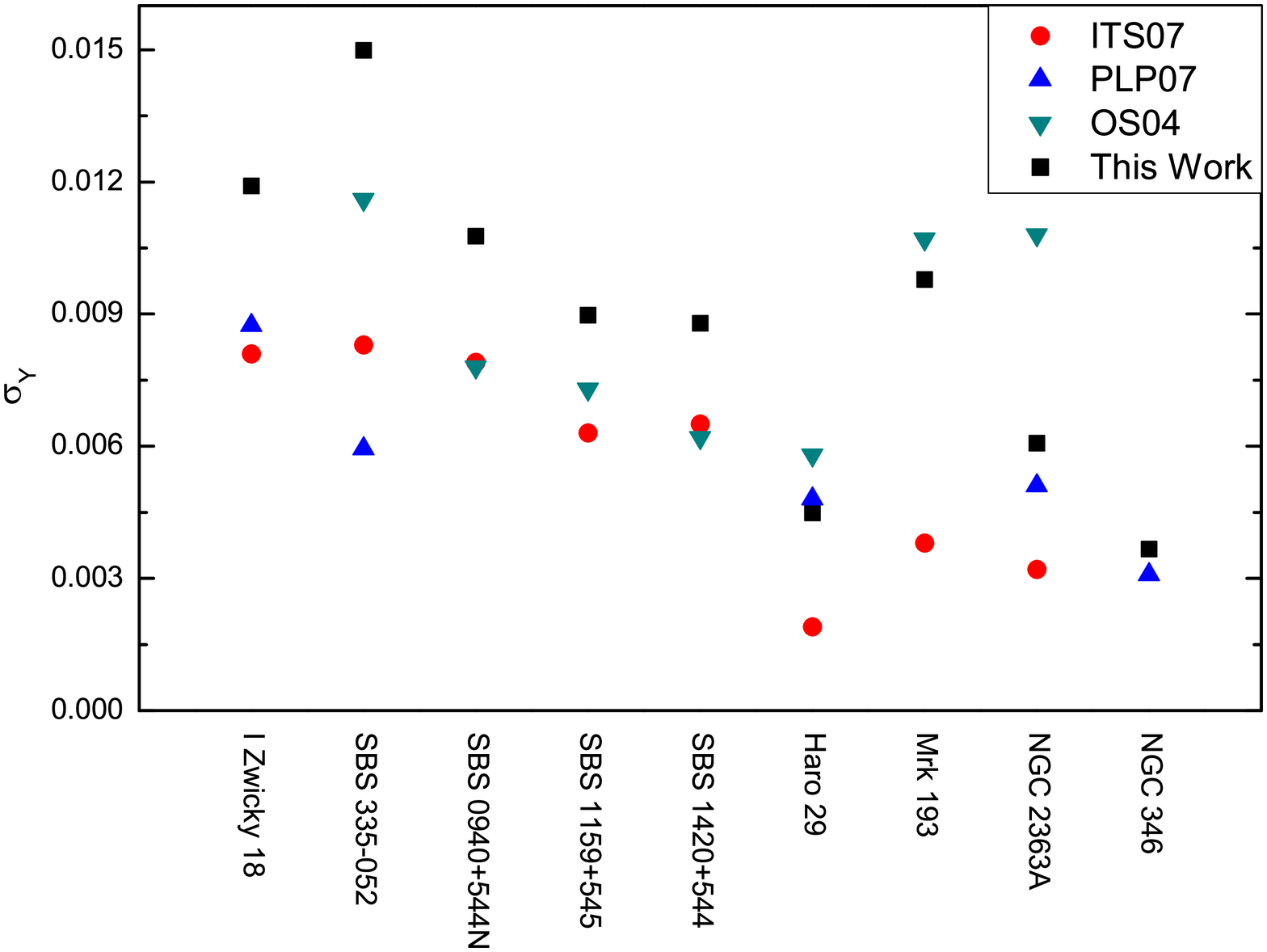}}
\caption{
A comparison of the calculated uncertainties on the individual data points from 
three different studies \citep[][and this work]{os04, its07, plp07}.
The data points are ordered from lowest oxygen abundance to highest oxygen 
abundance.
Note the trend of decreasing uncertainty  with increasing abundance.
Typically our uncertainties are larger than other previous analyses of the
same spectra.
}
\label{y_errs}
\end{figure}         
                                                                                                                                                                   
Since the main goal of this paper is an improved assessment of the errors on
the helium abundances derived from spectra of metal poor H~II regions, it is 
useful to compare the sizes of the error bars determined in this study with those
determined from the same spectra in other studies.
Figure \ref{y_errs} shows the sizes of the total errors associated with
each helium abundance determination for all of the objects studied here.
One surprising result of this comparison is that the relative size of the
errors is a strong function of oxygen abundance for {\it all} studies.
To our knowledge, this has never been seen or discussed before.
One possible explanation is that the most metal poor objects with suitably
bright H~II regions are relatively rare, so that the quality of the 
spectra for these objects may be lower.
Note also that typically the errors that we derive for each spectrum are
larger than previously derived by others.  The errors derived by ITS07 are
typically the smallest, and this is directly attributable to their method of
deriving the electron temperature which limits that value to a relatively
small range.  PLP07 derive electron temperatures in a similar method to
ours, but do not include a Monte Carlo analysis of the errors.  Thus, the
errors of PLP07 are typically intermediate to those of ITS07 and those 
presented here.

\begin{landscape}
\begin{deluxetable}{lccccccccc}
\tabletypesize{\scriptsize}
\tablewidth{0pt}
\tablecaption{Comparison of Physical Conditions and He$^+$/H$^+$ Abundance Solutions}
\tablehead{
\colhead{Object} &
\colhead{He$^+$/H$^+$}     &
\colhead{T$_e$}     &
\colhead{N$_e$}     &
\colhead{ABS(He~I)}     &
\colhead{$\tau$}     &
\colhead{C(H$\beta$)}     &
\colhead{ABS(H~I)}     &
\colhead{$\xi$ $\times$ 10$^4$}}
\startdata
\hline
&&&& \citetalias{os04} \\
\hline
SBS~0335-052 & 0.0763 $\pm$ 0.0049 & 15,940 $\pm$ 2710 &    347 $\pm$    942 &  0.1 $\pm$  0.2 &  4.6 $\pm$  0.9 &  0.121 $\pm$  0.014 &  1.3 $\pm$  1.9 & - &  \\
SBS~0940+544N & 0.0841 $\pm$ 0.0035 & 19,260 $\pm$ 2480 &    35 $\pm$    140 &  0.1 $\pm$  0.3 &  0.4 $\pm$  0.4 &  0.048 $\pm$  0.023 &  0.0 $\pm$  1.4 & - &  \\
SBS~1159+545 & 0.0838 $\pm$ 0.0031 & 19,330 $\pm$ 2100 &    75 $\pm$    116 &  0.0 $\pm$  0.1 &  0.5 $\pm$  0.4 &  0.065 $\pm$  0.016 &  0.2 $\pm$  1.1 & - &  \\
SBS~1420+544 & 0.0854 $\pm$ 0.0026 & 20,370 $\pm$ 1740 &     27 $\pm$    83 &  0.12 $\pm$  0.18 &  3.32 $\pm$  0.35 &  0.16 $\pm$  0.03 &  0.4 $\pm$  0.9 &  - &  \\
Haro~29 & 0.0844 $\pm$ 0.0024 & 16,680 $\pm$ 890 &     33 $\pm$     97 &  0.5 $\pm$  0.2 &  0.4 $\pm$  0.2 &  0.001 $\pm$  0.004 &  2.7 $\pm$  0.3 & - &  \\
Mrk~193 & 0.0845 $\pm$ 0.0023 & 15,643 $\pm$ 1130 &     23 $\pm$     95 &  0.37 $\pm$  0.14 &  2.29 $\pm$  0.35 &  0.254 $\pm$  0.019 &  0.36 $\pm$  0.51 & - &  \\
NGC~2363A & 0.0801 $\pm$ 0.0046 & 14,040 $\pm$ 1060 &    285 $\pm$    343 &  0.35 $\pm$  0.28 &  1.4 $\pm$  0.4 &  0.117 $\pm$  0.003 &  1.4 $\pm$  0.3 & - &  \\
NGC~346 & 0.0828 $\pm$ 0.0008 & 13,420 $\pm$  280 &     2 $\pm$     19 &  0.09 $\pm$  0.09 &  0.01 $\pm$  0.04 &  0.174 $\pm$  0.008 &  0.0 $\pm$  0.5 & - &  \\
\hline
&&&& Re-analysis \\
\hline
I~Zw~18 & 0.08008 $\pm$ 0.00469 & 17,588 $\pm$ 2951 &     76.7 $\pm$    277.2 &  0.42 $\pm$  0.36 &  0.42 $\pm$  0.56 &  0.00 $\pm$  0.00 &  4.76 $\pm$  0.82 &      7.89 $\pm$     70.34 &  \\
SBS~0335-052 & 0.08536 $\pm$ 0.00644 & 15,822 $\pm$ 2395 &    120.8 $\pm$    254.4 &  0.39 $\pm$  0.31 &  5.32 $\pm$  0.87 &  0.05 $\pm$  0.04 &  1.82 $\pm$  1.04 &    117.44 $\pm$    262.19 &  \\
SBS~0940+544N & 0.08506 $\pm$ 0.00484 & 16,689 $\pm$ 2124 &    149.3 $\pm$    277.0 &  0.09 $\pm$  0.20 &  0.61 $\pm$  0.56 &  0.00 $\pm$  0.00 &  0.03 $\pm$  0.17 &     25.82 $\pm$     33.46 &  \\
SBS~1159+545 & 0.08332 $\pm$ 0.00390 & 16,177 $\pm$ 2318 &    325.2 $\pm$    330.0 &  0.03 $\pm$  0.06 &  0.39 $\pm$  0.43 &  0.02 $\pm$  0.02 &  0.59 $\pm$  0.63 &     68.26 $\pm$    152.42 &  \\
SBS~1420+544 & 0.08939 $\pm$ 0.00393 & 20,214 $\pm$ 2695 &     83.1 $\pm$    234.0 &  0.18 $\pm$  0.16 &  2.89 $\pm$  0.62 &  0.16 $\pm$  0.03 &  0.02 $\pm$  0.14 &     30.30 $\pm$    200.16 &  \\
Haro~29 & 0.08576 $\pm$ 0.00174 & 17,092 $\pm$  665 &     37.1 $\pm$     48.3 &  0.52 $\pm$  0.15 &  0.13 $\pm$  0.16 &  0.00 $\pm$  0.00 &  3.38 $\pm$  0.24 &      0.20 $\pm$      0.39 &  \\
Mrk~193 & 0.08660 $\pm$ 0.00435 & 13,924 $\pm$ 2219 &    117.1 $\pm$    319.9 &  0.45 $\pm$  0.19 &  2.63 $\pm$  0.78 &  0.21 $\pm$  0.04 &  0.64 $\pm$  0.74 &   1003.06 $\pm$   8465.55 &  \\
NGC~2363A & 0.08328 $\pm$ 0.00244 & 13,578 $\pm$  992 &    194.2 $\pm$    203.1 &  0.46 $\pm$  0.18 &  1.55 $\pm$  0.34 &  0.12 $\pm$  0.01 &  0.29 $\pm$  0.44 &      5.20 $\pm$     49.25 &  \\
NGC~346 & 0.08777 $\pm$ 0.00168 & 12,847 $\pm$  708 &     30.2 $\pm$     70.2 &  0.37 $\pm$  0.15 &  0.03 $\pm$  0.09 &  0.13 $\pm$  0.01 &  0.10 $\pm$  0.28 &    358.72 $\pm$    758.49 &  \\
\enddata
\label{table:GTO}
\end{deluxetable}
\end{landscape}

\begin{deluxetable}{lcccccc}
\tabletypesize{\scriptsize}
\tablewidth{0pt}
\tablecaption{Comparison with ITS07 \& PLP07}
\tablehead{
\colhead{Analysis} &
\colhead{Y}     &
\colhead{T$_e$}     &
\colhead{N$_e$}     &
\colhead{$\tau$}     &
\colhead{C(H$\beta$)} &
\colhead{O/H}}
\startdata
\multicolumn{6}{l}{I~Zw~18} \\
\hline
ITS07		& 0.2485 $\pm$ 0.0081 & 18,100 $\pm$ 500	& 11  $^{+76}_{-0}$   & 1.30 $^{+0.00}_{-0.99}$ & & 0.16 $\pm$ 0.01 \\
		& 0.2415 $\pm$ 0.0104 & 18,600 $\pm$ 600	& 132 $^{+134}_{-75}$ & 0.42 $^{+0.51}_{-0.25}$ & & 0.15 $\pm$ 0.01 \\
PLP07		& 0.2505 $\pm$ 0.0087 & & 90  $\pm$ 80 & 0.06 $\pm$ 0.05	& 0.02 $\pm$ 0.01 & 0.22 $\pm$ 0.03 \\
This Work	& 0.2480 $\pm$ 0.0119 & 17,588 $\pm$ 2951 &     77 $^{+277}_{-77}$    &  0.42 $\pm$  0.56 &  0.00 $\pm$  0.00 & 0.15 $\pm$ 0.01 \\
\hline
\multicolumn{6}{l}{SBS~0335-052} \\
\hline
ITS07		& 0.2551 $\pm$ 0.0083 & 19,100 $\pm$ 600	& 12  $^{+42}_{-1}$   & 4.99 $^{+0.01}_{-0.44}$ & & 0.22 $\pm$ 0.01 \\
		& 0.2504 $\pm$ 0.0045 & 19,300 $\pm$ 400	& 127 $^{+39}_{-60}$  & 4.67 $^{+0.19}_{-0.17}$ & & 0.22 $\pm$ 0.01 \\
		& 0.2476 $\pm$ 0.0064 & 19,700 $\pm$ 600	& 176 $^{+51}_{-37}$  & 2.62 $^{+0.15}_{-0.31}$ & & 0.22 $\pm$ 0.01 \\
		& 0.2470  $\pm$ 0.0062 & 19,200 $\pm$ 600	& 325 $^{+84}_{-33}$  & 3.17 $^{+0.02}_{-0.63}$ & & 0.23 $\pm$ 0.01 \\
PLP07		& 0.2533 $\pm$ 0.0059 & & 329 $\pm$ 61 & 2.56 $\pm$ 0.35	& 0.03 $\pm$ 0.01 & 0.33 $\pm$ 0.03 \\
This Work	& 0.2595 $\pm$ 0.0150 & 15,822 $\pm$ 2395 &    121 $^{+254}_{-121}$   &  5.32 $\pm$  0.87 &  0.05 $\pm$  0.04 & 0.19 $\pm$ 0.01 \\
\hline
\multicolumn{6}{l}{SBS~0940+544N} \\
\hline
ITS07		& 0.2543 $\pm$ 0.0079 & 19,400 $\pm$ 300	& 104 $^{+79}_{-59}$  & 0.01 $^{+0.50}_{-0.00}$ & & 0.28 $\pm$ 0.01 \\
		& 0.2515 $\pm$ 0.0036 & 17,800 $\pm$ 500	& 10  $^{+23}_{-0}$   & 0.25 $^{+0.08}_{-0.16}$ & & 0.35 $\pm$ 0.01 \\
This Work 	& 0.2537 $\pm$ 0.0108 & 16,689 $\pm$ 2124 &    149 $^{+277}_{-149}$   &  0.61 $\pm$  0.56 &  0.00 $\pm$  0.00 & 0.26 $\pm$ 0.01 \\
\hline
\multicolumn{6}{l}{SBS~1159+545} \\
\hline
ITS07		& 0.2613 $\pm$ 0.0063 & 17,500 $\pm$ 400	& 77  $^{+63}_{-42}$  & 0.55 $^{+0.45}_{-0.24}$ & & 0.35 $\pm$ 0.01 \\
This Work 	& 0.2520 $\pm$ 0.0090 & 16,177 $\pm$ 2318 &    325 $^{+330}_{-325}$   &  0.39 $\pm$  0.43 &  0.02 $\pm$  0.02 & 0.30 $\pm$ 0.01 \\
\hline
\multicolumn{6}{l}{SBS~1420+544} \\
\hline
ITS07		& 0.2590  $\pm$ 0.0065 & 17,500 $\pm$ 500	& 91  $^{+86}_{-52}$  & 3.63 $^{+0.36}_{-0.27}$ & & 0.58 $\pm$ 0.02 \\
This Work 	& 0.2655 $\pm$ 0.0088 & 20,214 $\pm$ 2695 &     83 $^{+234}_{-83}$    &  2.89 $\pm$  0.62 &  0.16 $\pm$  0.03 & 0.54 $\pm$ 0.01  \\
\hline
\multicolumn{6}{l}{Haro~29} \\
\hline
ITS07		& 0.2546 $\pm$ 0.019  & 15,300 $\pm$ 900	& 12  $^{+258}_{-1}$  & 0.61 $^{+2.53}_{-0.34}$ & & 0.68 $\pm$ 0.01 \\
PLP07		& 0.2535 $\pm$ 0.0048 & & 61  $\pm$ 50 & 1.28 $\pm$ 0.30	& 0.02 $\pm$ 0.01 & 0.89 $\pm$ 0.10 \\
This Work 	& 0.2575 $\pm$ 0.0045 & 17,092 $\pm$  665 &     37 $^{+48}_{-37}$     &  0.13 $\pm$  0.16 &  0.00 $\pm$  0.00 & 0.58 $\pm$ 0.01 \\
\hline
\multicolumn{6}{l}{Mrk~193} \\
\hline
ITS07		& 0.2537 $\pm$ 0.0038 & 14,700 $\pm$ 100	& 11  $^{+19}_{-1}$   & 2.97 $^{+0.07}_{-0.38}$ & & 0.86 $\pm$ 0.02 \\
This Work       & 0.2592 $\pm$ 0.0098 & 13,924 $\pm$ 2219 &    117 $^{+320}_{-117}$    &  2.63 $\pm$  0.78 &  0.21 $\pm$  0.04 & 0.61 $\pm$ 0.02 \\
\hline
\multicolumn{6}{l}{NGC~2363A} \\
\hline
ITS07		& 0.2558 $\pm$ 0.0032 & 14,900 $\pm$ 400	& 14  $^{+73}_{-2}$   & 2.00 $^{+0.00}_{-0.48}$ & & 0.82 $\pm$ 0.02 \\
		& 0.2492 $\pm$ 0.0089 & 13,400 $\pm$ 300	& 13  $^{+127}_{-2}$  & 0.01 $^{+0.33}_{-0.00}$ & & 0.85 $\pm$ 0.05 \\
PLP07		& 0.2518 $\pm$ 0.0051 & & 285 $\pm$ 92 & 1.04 $\pm$ 0.40	& 0.01 $\pm$ 0.01 & 1.07 $\pm$ 0.08 \\
This Work 	& 0.2522 $\pm$ 0.0061 & 13,578 $\pm$  992 &    194 $^{+203}_{-194}$   &  1.55 $\pm$  0.34 &  0.12 $\pm$  0.01 & 0.69 $\pm$ 0.01 \\
\hline
\multicolumn{6}{l}{NGC~346} \\
\hline
PLP07		& 0.2507 $\pm$ 0.0031 & & 58  $\pm$ 33 & 0.10 $\pm$ 0.20	  & 0.00 $\pm$ 0.00 & 1.35 $\pm$ 0.20 \\
This Work 	& 0.2593 $\pm$ 0.0037 & 12,847 $\pm$  708 &     30 $^{+70}_{-30}$ &  0.03 $\pm$  0.09 &  0.13 $\pm$  0.01 & 1.02 $\pm$ 0.02\\
\enddata
\label{table:ITS_PLP}
\end{deluxetable}

\begin{deluxetable}{lcccc}
\tabletypesize{\footnotesize}
\tablewidth{0pt}
\tablecaption{Primordial Helium Regression Values}
\tablehead{
\colhead{Object} &
\colhead{He$^+$/H$^+$} &
\colhead{He$^{++}$/H$^+$} &
\colhead{Y} &
\colhead{O/H $\times$ 10$^4$}}
\startdata
I~Zw~18 & 	0.08008	$\pm$ 0.00469 & 0.0024 $\pm$ 0.0024 & 0.2480 $\pm$ 0.0119 & 0.15 $\pm$ 0.01 \\
SBS~0335-052 & 	0.08536	$\pm$ 0.00644 & 0.0023 $\pm$ 0.0023 & 0.2595 $\pm$ 0.0150 & 0.19 $\pm$ 0.01 \\ 
SBS~0940+544N & 0.08506	$\pm$ 0.00484 & 0.0000 		    & 0.2537 $\pm$ 0.0108 & 0.26 $\pm$ 0.01 \\ 
SBS~1159+545 & 	0.08332	$\pm$ 0.00390 & 0.0010 $\pm$ 0.0010 & 0.2520 $\pm$ 0.0090 & 0.30 $\pm$ 0.01 \\ 
SBS~1420+544 & 	0.08939	$\pm$ 0.00393 & 0.0011 $\pm$ 0.0011 & 0.2655 $\pm$ 0.0088 & 0.54 $\pm$ 0.01 \\ 
Haro~29 & 	0.08576	$\pm$ 0.00174 & 0.0011 $\pm$ 0.0011 & 0.2575 $\pm$ 0.0045 & 0.58 $\pm$ 0.01 \\ 
Mrk~193 &       0.08660 $\pm$ 0.00435 & 0.0010 $\pm$ 0.0010 & 0.2592 $\pm$ 0.0098 & 0.61 $\pm$ 0.02 \\
NGC~2363A & 	0.08328	$\pm$ 0.00244 & 0.0012 $\pm$ 0.0012 & 0.2522 $\pm$ 0.0061 & 0.69 $\pm$ 0.01 \\
NGC~346 & 	0.08777	$\pm$ 0.00168 & 0.0000 		    & 0.2593 $\pm$ 0.0037 & 1.02 $\pm$ 0.02 \\
\enddata
\label{table:PH}
\end{deluxetable}

\section{Discussion \& Future Improvements} \label{Discussion}

The preceding work  in Papers I and II emphasized the importance of the systematic uncertainties in calculating y$^{+}$ and the need for Monte Carlo analysis.  Extending that work to allow for self-consistent analysis reflecting the common temperature and reddening parameters between the hydrogen and helium lines does not diminish the importance of the systematic uncertainties or the case for the larger errors determined via Monte Carlo.  In fact, the addition of the neutral hydrogen collisional emission, introduced with a new physical parameter, the neutral hydrogen fraction, highlights that abundance errors may be even larger than previously thought.  In addition to the need for observations to constrain systematic uncertainty as much as possible, sections \ref{Updated Emissivities} \& \ref{Integrated} both demonstrate the significant impact that relatively small systematic errors in the observed fluxes, even of the weakest lines, can have on y$^{+}$.  Clearly high quality spectra are critical for meaningful analysis.

Given higher resolution spectra, the emission line fluxes could be measured 
separately from the effects of underlying absorption.  This is possible 
due to collisional broadening of the stellar absorption lines, such that 
the emission line can be measured relative to its local continuum.  
In other words, the absorption effect in the continuum would be visible in
depth and extent with a narrow emission line at the minimum.  As a result 
the flux could be measured relative to the absorption feature rather than 
the broader continuum.  The underlying helium and hydrogen (through $H\beta$) 
lines both directly impact the abundance determination. Furthermore, decreasing 
the number of solution parameters (by eliminating the parameters associated 
with underlying absorption) will in general decrease the uncertainty.  Therefore, 
an improvement in the abundance determination is ensured.  Typically, underlying
Balmer emission is much easier to detect in the spectrum of a metal poor H~II 
region than underlying He~I emission, so starting with eliminating the 
correction for underlying Balmer absorption is the most promising.

Though difficult to gauge theoretically,  some insight into the degree of improvement  
from more accurate measurements can be gained through synthetic data.  
Assuming a typical value of 1.0 \AA~for the underlying Hydrogen and Helium absorption, 
we can generate corresponding fluxes when the remaining parameters are chosen.  
For  $\tau = 1$, C(H$\beta$) = 0.1, $T$ = 18000 K, $\xi = 10^{-4}$ , and $y^{+}$ = 0.08, 
we generated fluxes to determine the helium abundance
when either the underlying absorption is assumed to be known or unknown.
At a low density of 10 cm$^{-3}$ there was no improvement in the determination of $y^+$ due to a large bias in the Monte Carlo stemming from the requirement that the density remain positive.  At a more characteristic density near 100 cm$^{-3}$ the benefit was significant.  The returned value for the helium abundance was y$^{+}$ = 0.080997 $\pm$ 0.003735 and y$^{+}$ = 0.080002 $\pm$ 0.002841 for the underlying absorption solved and removed cases respectively.

Ultimately, the sensitivity of the determination of the primordial helium abundance demands that analysis be self-consistent and comprehensive; unfortunately efforts in providing such an analysis yield uncertainties that limit the utility of that determination.  Nevertheless, fully understanding the terrain of potential obstacles is necessary to direct further measurement efforts so as to provide a reliable calculation of the primordial helium abundance.  Apart from improving the spectra, the neutral hydrogen collisional correction for the weaker hydrogen lines can be easily improved from extended principal quantum number collisional strengths.  Furthermore the use of a further hydrogen line would benefit the neutral hydrogen fraction and hydrogen absorption since those two parameters are determined exclusively through the three hydrogen lines.  
The next Balmer line, H$\epsilon$ ($\lambda$3970) is blended with Ne~III ($\lambda$3967).  The contribution of Ne~III $\lambda$3967 to the H$\epsilon$ $\lambda$3970 could 
be corrected for by scaling from the stronger Ne~III $\lambda$ 3868 line, or by observing at sufficiently higher spectral resolution in order to deblend the two lines.
H8 ($\lambda$3889), is blended with He~I $\lambda$3889, and would not be helpful independent for this type of analysis.  
This leaves the relatively very weak H9 ($\lambda$3835) and H10 ($\lambda$3798) as the best candidates.   
The use of other emission lines which are sensitive to the density may also be useful in reducing the uncertainty due to the degeneracy between density and temperature.  
However, most candidates with the appropriate electronic structure are not the dominant ionic species in metal-poor H~II regions (e.g., [S~II], [O~II]).
A possible exception is the [Cl~III] emission line pair $\lambda\lambda$5518, 5538.


In summary, it should be emphasized that even though the precision of the primordial helium abundance is not increased, the benefit of the analysis improvements is precisely in demonstrating the full effect of the systematic errors in restricting the abundance determination.

Note Added:  On completion of this paper, we became aware of the newest result of 
\cite{it10}.  While we find a very similar central value for $Y_p$, we disagree with their
conclusion calling for physics beyond the standard model of BBN.  The cumulative uncertainties
described above are sufficiently large that, within these uncertainties, standard BBN is not 
presently threatened on the basis of \he4 observations.

\acknowledgments

We would like to Yuri Izotov for providing additional information
concerning his spectra.
Preliminary results of this work were presented at the conference
IAU Symposium 268, ``Light Elements in the Universe,'' and we would
especially like to thank the conference organizers Corinne Charbonnel and 
Monica Tosi for the opportunity to do so. 
We would like to thank Y.\ Izotov, 
A.\ Peimbert, M.\ Peimbert, G.\ Steigman,
E.\ Terlevich, and R.\ Terlevich, for informative and
valuable discussions.
The work of KAO is supported in  part by DOE grant
DE-FG02-94ER-40823. EDS is grateful for partial support from 
the University of Minnesota.
This research has made use of NASA's Astrophysics Data System
Bibliographic Services and the NASA/IPAC Extragalactic Database
(NED), which is operated by the Jet Propulsion Laboratory, California
Institute of Technology, under contract with the National Aeronautics
and Space Administration.


\end{document}